\documentclass[twocolumn]{article}

\AtBeginDocument{%
  \providecommand\BibTeX{{%
    \normalfont B\kern-0.5em{\scshape i\kern-0.25em b}\kern-0.8em\TeX}}}

\usepackage{enumitem}
\usepackage{centernot}
\usepackage{graphicx}
\usepackage{hyperref}
\usepackage{amsfonts}
\usepackage[disable]{todonotes}
\usepackage{acronym}
\usepackage{multirow}
\usepackage{tikz}
\usepackage{amsmath}

\usepackage[normalem]{ulem}

\usepackage{titling}

\newif\iflong

\acrodef{OR}[OR]{Onion Routing}
 \acrodef{PPT}[PPT]{probabilistic polynomial time}
 
 \def\Ando{Dropping-Bound}
 \def\Hevia{Optimality-Bound}
 \def\Gelernter{Counting-Bound}
 \def\Trilemma{Trilemma} 
 
 \acrodef{ACN}{Anonymous Communication Network}
\acrodef{AC}{Anonymous Communication}
\begin{document}
\title{SoK on Performance Bounds in Anonymous Communication}

\author{Christiane Kuhn\\
\emph{Karlsruhe Institute of Technology}\\
\emph{christiane.kuhn@kit.edu}
\and
Friederike Kitzing\\
\emph{TU Dresden}\\
\emph{friederike.kitzing@mailbox.tu-dresden.de} 
\and
Thorsten Strufe\\
\emph{Karlsruhe Institute of Technology}\\
\emph{strufe@kit.edu}
}

%


\maketitle

\subsection*{\emph{Errata}}\emph{This work was originally published at WPES 2020 \cite{kuhn2020sok}.}  \emph{Contrary to the original version's claim, the Dropping-Bound needs to corrupt the receiver to be able to distinguish the real message from dummy messages, which could be applied link-based. Further, Figure \ref{fig:triangle} contained a shift in the axis labeling. Both is corrected in this version.}

\begin{abstract}
 Communicating anonymously comes at a cost -- and large communities have been in a constant tug-of-war between the development of faster protocols, and the improvement of security analyses. Thereby more intricate privacy goals emerged and more detailed bounds on the minimum overhead necessary to achieve them were proven.
 The entanglement of requirements, scenarios, and protocols complicates analysis, and the published results are hardly comparable, due to deviating, yet specific choices of assumptions and goals (some explicit, most implicit).

 In this paper, we systematize the field by harmonizing the models, comparing the proven performance bounds, and contextualizing these theoretical results in a broad set of proposed and implemented systems.
By identifying inaccuracies, we demonstrate that the attacks, on which the results are based, indeed break much weaker privacy goals than postulated, and tighten the bounds along the way.  We further show the equivalence of two seemingly alternative bounds. Finally, we argue how several assumptions and requirements of the papers likely are of limited applicability in reality and suggest relaxations for future work.
\end{abstract}

\section{Introduction} \label{sec:introduction}
 \acp{ACN} have been developed, and their underlying concepts and properties have been investigated throughout the last 30 years.
Improvements to the protocols often aim at better performance, but are also guided by progressively sophisticated attacks. Those improvements  are still of utmost importance. Millions of users\footnote{cf. \url{https://metrics.torproject.org/userstats-relay-country.html} } rely on the protection of Tor~\cite{dingledine04tor}, even though attacks are known~\cite{murdoch2005low, bauer2007low, abbott2007browser}.

Identifying fundamental limits of this trade-off between performance and privacy can greatly aid developers in their design of new protocols. 
It is, however, a challenging task, as the analyzed system, the domain of possible adversaries, and even the definition of anonymity in itself are complex.

All existing formal analyses concluded that a prohibitively high overhead is necessary to achieve provable anonymity. 
This consistently pessimistic message does not help the developers of \acp{ACN} much.
To really understand, assess, and make use of the bounds on the efficiency of provable \acp{ACN}, the complex theoretical proofs have to be investigated in depth, compared, and considered from a practical viewpoint.

With this paper we help to close the gap between the theoretical proofs and their practical ramifications.
The corresponding papers frequently are very technical and mostly missing practical perspectives.
They employ various notations, define different models, and address diverse privacy goals (although all are called \emph{anonymity}) 
under a large variety of assumptions on the users, their behavior, and adversary models.
To understand the performance of ACNs, we need to show which claim on necessary overheads comes with which choice of requirements and assumptions.

Thus, we systematize the underlying properties of the existing analyses and make their results more accessible.
This requires to examine the assumed adversary models, and the privacy goals that are explicitly claimed, as well as those that actually are analyzed, given several implicit restrictions. 
Implicit assumptions on the protocols and sending behavior have to be made explicit, to allow for comparison.

In the last part of this process, we finally investigate and compare all bounds, as proven in the papers.
They all break certain privacy goals, and prove the minimum overhead that is necessary to prevent the considered attack.
Our research initially reveals, that the situation actually is worse, than the papers proclaim:
The presented attacks indeed break privacy goals that are much weaker than what the studies target.
We hence tighten the derived bounds, and, in one case, discover that the necessary overhead is even higher than concluded by the authors. Further, we discover the equivalence of two bounds, which have different perspectives on the domain.

We put those conclusions into perspective, as we contextualize the given notions and bounds within a broad field of actual and proposed anonymous communication networks.

Finally, we discuss how specific details of the models and assumptions, which are chosen either to simplify analysis or to express theoretical worst cases, are causing the minimum overhead to be very high.
Extracting these peculiarities, we identify the challenges for future research on bounds: more realistic user behavior and  relaxed assumptions.

In brief, this paper provides the following contributions:
\begin{itemize}
 \item setting out, tightening, harmonizing and comparing the fundamental concepts of the different bounds:
 \begin{itemize}
  \item the underlying attacks,
  \item the implicit privacy goals,
  \item the required adversarial capabilities,
  \item the deviating protocol assumptions,  and
  \item the derived bounds;
 \end{itemize}
  \item proving two seemingly different bounds equivalent,
 \item discussing implications on real protocols, and
 \item suggesting improvements for the formal analyses, to provide more applicable, and convincing insights on the actual cost of anonymity on the Internet.
\end{itemize}

\paragraph*{Outline} Section~\ref{sec:background} contains the background and Section~\ref{sec:comparison} an overview of the (tightened) bounds, while Section~\ref{comparisonREAL} compares the bounds in detail. Section~\ref{sec:implications} sheds light on their relation to proposed \acp{ACN} and Section~\ref{sec:practical} explains limitations from a practical viewpoint. Section~\ref{sec:conclusion} concludes the paper. Further, the Appendix contains formal details for the privacy definitions, the argumentation for the tightening of the bounds, tables to summarize the notation and results, technical parts of the proofs, information about receiver privacy and related results in slightly different research areas.

\section{Background} \label{sec:background}
In this section, we first introduce the ACN setting and explain the basic techniques used for building ACNs. Thereafter, we give a first, rough idea how bounds are shown. Further, we introduce the formalization needed for our extensive comparison: privacy goal definitions and  our notation.

\subsection{ACN Background}
While encryption is a well known measure to protect the content of messages, packet switched networks leak other properties that require protection. For example the sender and receiver of a communication should be hidden from profiling attempts of companies. \ac{AC} tries to solve this challenge. 
 

We discuss \ac{AC} in the setting of multiple unicast communications, like several client-server applications or messaging between users on the internet. 
Every communication is a message  that is sent from a sender, possibly forwarded by intermediate nodes, and finally received at the receiver. Such \acp{ACN} consist of nodes playing two roles; users (senders and receivers) and service providers (intermediate nodes). 
Participants in some systems play both roles for different communications.
We distinguish between the  \emph{integrated system model}, where the receiver is part of the \ac{ACN} and acts according to the \ac{ACN} protocol, and the \emph{service model}, where messages are anonymized as a service and the receiver, e.g. webserver, can be completely unaware of the \ac{ACN}.

Depending on the use case, the privacy goals and assumed adversaries differ. 
A common privacy goal is to hide some behavior of a sender, e.g. who sent a certain message, or how often a sender sends. 
This is usually achieved by hiding the sender among others. 
This set of users who are possible senders is called the \emph{anonymity set}. 
Other than protecting the sender it can be a goal to protect the receiver or sender-receiver relationships.

Typical adversary models assume a global passive adversary, who eavesdrops on all links, or constrained versions limited to only a subset of links.
Passively corrupted receivers, or intermediate nodes, which additionally leak their keys to the adversary, are a common extension. 
Stronger models even allow the adversary to modify, drop, insert and delay packets at the controlled parts of the network.

\subsection{ACN Techniques}
We explain the conceptual ideas to achieve a privacy goal roughly and refer the reader to \cite{edman2009anonymity, shirazi2018survey} for detailed surveys.

\subsubsection{Indirection}
\emph{Onion routing}~\cite{goldschlag1996hiding} and \emph{mix networks}~\cite{chaum1981untraceable} hide which sender has sent a certain message to whom, by relaying it over multiple hops. 
Applying layered encryption or shuffling of messages (in the case of mix networks), they ensure unlinkablity of incoming and outgoing messages at honest intermediate nodes. 
They protect against passive adversaries that corrupt some receivers and intermediate nodes. 
Extensions provide protection against  stronger adversaries.

\subsubsection{Superposition} \label{sec:backDC}
\emph{DC-Nets}~\cite{chaum1988dining} implement superposition to broadcast one message per round without leaking which of the participants is the sender, as any participant is sending a part necessary to recover the message.  
However, sending more than one message per round leads to collisions of messages and none of them are interpretable. 
Thus, a collision avoidance scheme is usually assumed.

\emph{Private information retrieval} (PIR)~\cite{chor1995private} allows to request and deliver an entry of a database without disclosing which entry was requested. 
Using e.g. superposed shares of data, it allows a receiver to anonymously request messages that are stored at a database, thus protecting the recipient's privacy. 
Some approaches reverse the idea to protect the senders. 

\subsubsection{Dummy messages}
\emph{Dummy messages} do not transmit useful information. They instead are sent to hide sending of ``{\em real messages}'', which contain useful information for a receiver.
Dummy messages are sent randomly, or systematically according to user synchronization, and later dropped by some part of the protocol, usually an intermediate node or the receiver. 

\subsection{Privacy Goals}\label{back:notions}

To sort the bounds on anonymity, we need a better understanding of what ``anonymity'' actually means in each case. Following the formal definitions of ~\cite{kuhn2018privacy}, we will distinguish the following forms of anonymity in this work:

\emph{Communication Unobservability} ($C\bar{O}$): Anything regarding the communications, even how many communications are happening, has to be hidden from the adversary. 

\emph{Receiver Unobservability} ($R\bar{O}$): Everything about the receivers, including any information about how many message they received, is hidden. The senders and their messages however can be learned by the adversary.

\emph{Sender Unobservability} ($S\bar{O}$): Everything about the senders, including any information about how many messages they sent, is hidden. The receivers and their messages however can be learned by the adversary.

\emph{Sender-Message Unlinkability}\footnote{$(SM)\bar{L}$ is called Pair-Sender-Message Unlinkability, $(SR)\bar{L}$:  Pair-Sender-Receiver Unlinkability in \cite{kuhn2018privacy}.} ($(SM)\bar{L}$): Only the fact that a message and its sender belong together is hidden. Therefore, for any two (honest) senders, even if the adversary knows that one of these suspects sent a certain message, she cannot tell which of the two senders it was. Besides many other things, this allows that the adversary learns which sender sends how many messages and to whom each sender is communicating. 

\emph{Sender-Receiver Unlinkability}\footnotemark[2]\todo{check footnote} ($(SR)\bar{L}$): Only the fact that a sender and receiver communicate with each other is hidden. Therefore, for any two (honest) senders, even if the adversary knows that one of these suspects communicates with a certain receiver, she cannot tell which of the two senders it is. Besides many other things, this allows that the adversary  learns which sender sends how often and which sender sends which messages. 

We refer the interested reader to Appendix~\ref{app:privacyNotions} and \cite{kuhn2018privacy} for a formal definition and note that all our comparisons and improvements work similarly on the underlying formal model.

\section{Bounds Overview}\label{sec:comparison} \label{sec:landscape}

 To increase the privacy of otherwise unprotected communication, ACN techniques necessarily create overhead. 
The dominating strategy to prove that a minimum amount of overhead is needed to achieve a privacy goal is based on attacks:
According to assumptions and protocol requirements the attack is argued to succeed, unless the protocol creates a certain amount of overhead.

We consider the protocol assumptions, privacy goal, adversary model,  the attack idea and the derived performance bound as fundamental details of each bound.


Analyzing the proofs in the reports, we realized that their minimum amount of overhead is already necessary to achieve much weaker privacy goals for weaker adversaries than claimed in the works, and hence we tightened the bounds (and in one case correct the necessary overhead).
Appendix~\ref{sec:worse} describes this analysis in detail. Here we mention the improvements only briefly and then use the improved results throughout the rest of this paper.

Further, we limit ourselves to explain the sender goal based bounds in the main part and refer the interested reader to Appendix~\ref{sec:receiver} for a discussion of receiver goals and to Appendix~\ref{sec:discussion} for distantly related considerations on overhead.

In this section, we give a high-level overview of the bounds in order of increasing strength of privacy notions that they actually relate to. Their details are discussed as part of the comparison in the next section.  


\subsection{\Ando\ \cite{ando2019complexity}}

We call this bound ``\Ando'' because the attack relies on dropping packets.

\subsubsection*{Protocol Assumptions}
The bound only considers onion routing and mix networks. It relies on the implicit assumption that messages are successfully delivered with high probability.

\subsubsection*{Privacy Goal: $(SR)\bar{L}$} The report analyses for the strongest possible goal $C\bar{O}$.
 The bound, however, already applies for one of the weakest notions, $(SR)\bar{L}$.
It defines that for any two (honest) senders the adversary must not learn which of them communicated with which of two receivers. Except this, she can learn anything, including e.g. how often each sender sends. Note that she can especially learn the fact that both candidate senders communicated with one of the two receivers, but not who communicated with whom.

\subsubsection*{Adversary Model}
The paper states the assumption of active adversaries. 
Note, that the only necessary activity is to drop packets, though: 
The adversary can drop packets on the links of at least one sender and can observe at at least one receiver. Further, the adversary knows that this receiver expects a packet\footnote{This is due to the formal definition of the privacy goal. Practically, we can however understand this as external information the adversary gained, e.g. because the application requires a stream of messages.}.

\subsubsection*{Attack}
The adversary chooses a candidate sender, drops as many messages sent by this sender as she can, and observes whether an expected message still arrives at the receiver, or not.
She guesses her victim to be the real sender if no message arrives, and the alternative sender if it does.

\subsubsection*{Bound}
Preventing this attack requires some overhead, which we can measure in added bandwidth and latency.
Sending increasing numbers of redundant messages over alternative first hops requires higher bandwidth, but it improves the likelihood of delivery, as it reduces the chance 
that all paths start with adversarial links. 
Choosing longer paths increases latency but also the chance of an alternative message to be relayed through the victim sender and subsequently dropped by the adversary. 
This terminally reduces the accuracy of the adversary's guess\footnote{Note that this assumes an integrated system model, in which users also act as intermediate nodes.}.

The precise bound, which we discuss later, follows from calculating the adversary's advantage given an assumed cost.

\subsection{\Trilemma\ \cite{das2018anonymity}}

The ``\Trilemma'' bound claims that only two out of three desirable properties can be achieved in conjunction:
low bandwidth overhead, low delays, and strong\footnote{We show in App. ~\ref{sec:worseTrilemma} that it also holds for a 
weaker definition of provable anonymity.}
anonymity .

\subsubsection*{Protocol Assumptions}
The analysis assumes only a single receiver, and two suspect senders. All messages are delivered in at most $l_{max}$ rounds after sending. 

Further, it considers protocol to use a fixed amount of real and dummy messages per round. Two different user behaviors are specified:
In the \emph{synchronized model} one sender is assumed to send its real message and all other users synchronize to decide who sends dummy messages in this round. In the \emph{unsynchronized model}, any sender sends their real message in the current round with the fixed probability $p'$ and dummy messages with the fixed probability $\beta$. 

Although no restriction in the type of protocol is made explicit, we expect the bound to hold only for onion routing and mix networks, as at least one intermediate node is assumed.

\subsubsection*{Privacy Goal: $(SM)\bar{L}$}
While the report discusses $S\bar{O}$,  the \Trilemma\ already applies for one of the weakest notions $(SM)\bar{L}$. It defines that for any two (honest) senders the adversary cannot know which of them sent which message. Except this, the adversary can learn anything, including e.g. how often each sender sends. Note that she can especially learn the fact that both candidate senders sent a message, but not who sent which message.

\subsubsection*{Adversary Model}
The \Trilemma\ distinguishes two models: 

\emph{The ``non-compromising''\footnote{This name is used to distinguish it from the compromising adversary, even though the non-compromising adversary compromises the receiver.} adversary: }
The attacker controls the receiver and the links adjacent to the two suspected senders.
 
\emph{The compromising adversary}
The adversary additionally fully controls some intermediary nodes.

\subsubsection*{Attack}
The paper discusses two ways of identifying the real sender upon reception of a message at the corrupted receiver.

\emph{The non-compromising adversary:}
First, the attacker monitors the sending behavior of both suspected challenge users. 
If one user has not sent any message (real/dummy) within the 
$l_{max}$ rounds before the considered message is received, the other must be the sender.

\emph{The compromising adversary:}
In addition to the attack above, the adversary follows a second strategy:
With some probability she is able to observe all hops of either the challenge message, or the message sent by the alternative sender. 
She then can identify the sender-message pair and tell the sender of the challenge message.

\subsubsection*{Bound}
Increasing either latency or bandwidth helps preventing these attacks:
Sending dummy messages at higher probabilities translates to a  larger set of candidate users that might have sent the message, and thus a better chance that the alternative suspect is in it. 
Increasing the number of hops, and hence the latency, reduces the chance of all intermediate nodes being corrupt, and also increases the interval during which the message may have been sent, which again translates to a larger set of candidate senders. 

The precise bound follows from calculating the probabilities of the above mentioned events in which the adversary can unambiguously identify the sender, subject to the assumed bandwidth and latency overhead.

\subsection{\Gelernter\ \cite{gelernter2013limits}}

The ``\Gelernter''\ relies on counting delivered packets.

\subsubsection*{Protocol Assumptions}
 None.
 
\subsubsection*{Privacy Goal: $S\bar{O}$}
No information about any sender can leak. This includes for example that even if someone sent all messages, the adversary does not know whether or not she sent any message at all.

\subsubsection*{Adversary Model}
The honest, but curious adversary corrupts all receivers and the links of at least one honest sender.

\subsubsection*{Attack}
The \Gelernter's privacy goal implies that all participating senders could have sent \emph{all} real messages. 
The attacker now attempts to exclude at least one of them, by counting the number of messages they are sending.
Knowing the number of real messages that are received (as the adversary controls the receivers), the adversary can exclude any sender who sent less messages. 

\subsubsection*{Bound}
The protocol cannot deliver more real messages to corrupt receivers  than any sender sends in real and dummy messages. 

%



\subsection{\Hevia\ \cite{hevia2008indistinguishability}}

We call this bound ``\Hevia'' because it is included in Hevia and Miccianchio's proof that their way of adding dummy messages is optimal from a performance point of view.

\subsubsection*{Protocol Assumptions}
All sent, real messages are delivered.

\subsubsection*{Privacy Goal: $S\bar{O}$} Alike the \Gelernter . 

\subsubsection*{Adversary Model}
The adversary observes the links of at least one honest sender and knows how many real messages will be sent in total\footnote{This is due to the formal definition of the privacy goal. For practical reasons, we might however also think of this as external information the adversary gained through another channel.}.

\subsubsection*{Attack}
The adversary again tries to infer that some user did not send all real messages. Therefore, she counts the number of messages each sender sends and concludes that this sender cannot have sent all, if the number is less than the total amount of messages.
 
\subsubsection*{Bound}
Each sender has to send as many (real and dummy) messages  as real messages will be sent.

\section{Comparison} \label{comparisonREAL}
We first compare the \Gelernter\ and \Hevia, to find that they only differ in small nuances. After that we compare the remaining bounds, aspect by  aspect.

\subsection{\Gelernter\ and \Hevia\ are equivalent}

Both bounds arise from the same argument: Considering a number of real messages that have been sent, 
anybody who sent less messages in total cannot have sent them all.

Protecting the privacy hence requires generating enough dummy messages to ensure that every sender sends as many times as real messages are delivered by the protocol.

While the privacy goal and resulting bound are identical (see Appendix~\ref{app:HeviaGelernter}), the authors of the two bounds looked at this from slightly different angles: 
The \Gelernter\ does not have any assumptions on the protocol, but instead requires that the receiver is corrupted, such that the adversary can count the delivered messages. 
The \Hevia\ however does not corrupt the receiver, but instead silently assumes that all messages are delivered and exploits the fact that the adversary knows how many real  messages are sent in total. Therefore, the adversary trivially also learns the number of delivered messages.

Both derive the same bound, but their conclusions differ correspondingly:
The \Gelernter\ limits the number of delivered messages, while the \Hevia\ requires the senders to send enough dummy messages.

We continue to use the \Gelernter\ as representative for both.

\subsection{Protocol Assumptions} \label{sec:assumptions}
The papers state, but also silently make assumptions regarding sending behavior, delivery guarantees, and supported protocols.

\subsubsection{Sending Behavior}

\label{sec: L}
The \Gelernter\ and \Ando\footnote{This is not to be confused with the assumptions for the protocol proposed in the same paper \cite{ando2019complexity}, where the messages are sent at the same point in time.} make no assumption about the distribution of sending events per round.
The \Trilemma\ however considers a specific sending behavior with fixed amounts dummy and real messages per round, and their synchronized and unsynchronized sending model.


\subsubsection{Delivery Guarantees}\label{sec: no guaranted delivery}

The \Gelernter\ does not consider a maximum delivery delay. As only $\frac{1}{n}$ of the sent messages (dummy and real) reach their destination, some messages might not be delivered.

The \Ando\ silently assumes successful message delivery. Missing messages otherwise could not be interpreted as successful attacks by the adversary, but they could be an artifact of the protocol.

The \Trilemma\ assumes  a maximum delay the network adds.
Note that in the synchronized setting this guarantees that all messages can successfully be sent and received, as users get assigned one of $n$ rounds to send their message into the network. 
The unsynchronized setting in contrast does not provide this kind of certainty, since a user can only send her message based on the result of a coin flip. Therefore, in every round there is some probability that a certain user has not been able to send their message yet (even though this probability is negligible after enough rounds).

\subsubsection{Protocol types}\label{sec:trilemmaNoLatency}
The \Gelernter\ applies to all types of ACN protocols, as it only considers the number of sent and received messages.
The \Ando, in contrast, only applies to onion routing and mix networks.

The \Trilemma\ states that no protocol with a minimal latency of $l_{max}=1$ can achieve their privacy goal, as the resulting advantages of their attack are  non-negligible\footnote{$\delta \geq \frac{1}{2}$ for the unsynchronized and $\delta \geq 1$ for the synchronized  setting (see Appendix \ref{app:latencyNull})}. 
However, there exist protocols with this minimal latency achieving even stronger privacy notions against the considered adversary model, like the secure multi-party computation protocol as discussed in \cite{gelernter2013limits} or the well-known DC-Net, which is proven to achieve a stronger notion than targeted by the \Trilemma\ in  \cite{gelernter2013limits}. 
The authors recognize this limitation in later work \cite{das2019not}, and we suspect their bound to apply only to ACNs following the onion routing or mix network paradigms.

 \subsection{Privacy Goals}\label{sec:mainNotions}
 Although all bounds claim to hold for the privacy goal ``anonymity'', the protection at which their overhead becomes necessary differs.

The \Gelernter\ targets $S\overline{O}$, the strongest goal of these analyses\footnote{It is called ``sender anonymity'' in  \cite{hevia2008indistinguishability} and is shown to map directly to $S\bar{O}$ in \cite{kuhn2018privacy}.}. 
It is a very strong notion that protects not only the linking of sender-message and sender-receiver pairs, but even the frequency of sending, which for instance after a critical event could jeopardize the sender's safety.

Both other bounds target weaker notions with no direct relation to each other. 
The \Trilemma\ considers $(SM)\bar{L}$, which only prevents linking sender-message pairs, while the notion of the \Ando\ only prevents linking sender-receiver pairs ($(SR)\bar{L}$).
Both allow the adversary to succeed in linking other properties, and allow to learn, for example, the number of real messages each user has sent.

\begin{figure}[h]
	\center
	\includegraphics[width=0.4\textwidth]{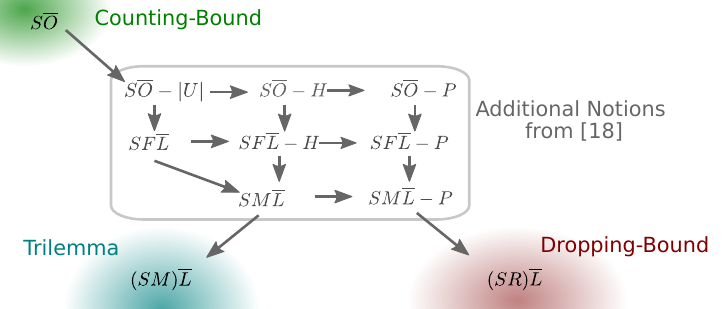}
	\caption{Excerpt of the hierarchy of~\cite{kuhn2018privacy} with the privacy goals of the bounds highlighted. Arrows point from stronger to strictly weaker goals.}
	\label{fig:NotionsFinal}
\end{figure}

To visualize the extent of the difference, note that the hierarchy provided by \cite{kuhn2018privacy} actually defines multiple other privacy goals, which do not further need in this work, in between the ones targeted by the different bounds  (see Figure~\ref{fig:NotionsFinal}).

\subsection{Adversary Models}\label{sec:adversaryComp}
The bounds rely on different adversary models, which we depict in Figure~\ref{fig:adversary} (the comparison to the models as stated  in the papers is provided in Appendix \ref{sec:worseAdv}).

\begin{figure}[h]
\begin{center}
\includegraphics[width=0.4\textwidth]{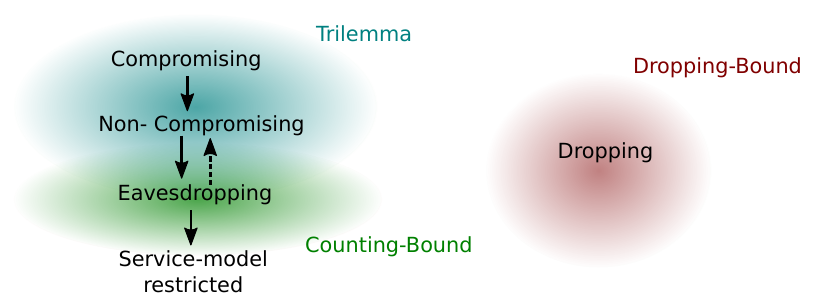} 
\caption{Hierarchy of adversary models. Hierarchically lower adversary model are weaker. 
The dotted arrow represents the additional relation caused by ignoring the  number of observed victims.} \label{fig:adversary}
\end{center}
\end{figure}



It is sufficient for all the attacks to compromise or influence outgoing links or the attached relays at the sender, as well as the receiver (or its incoming links/attached relays), as this enables to correlate events at the terminals of the communication.
The adversaries in the \emph{\Gelernter} and \emph{\Trilemma\ in the non-compromising case} are virtually identical: one that corrupts all links of the victim sender(s) and the receiver. 
Their only difference lies in the number of victims, as the \Trilemma\ considers two, and the \Gelernter\ only a single victim to be monitored.
The \emph{\Trilemma\ in the compromising case} additionally allows to passively compromise some intermediary protocol parties, i.e. to learn their keys and eavesdrop at them.

Only the {\em \Ando} allows the adversary to drop messages, and hence considers an active model.
Eavesdropping capabilities on the sender links are not strictly required, and it is at least conceivable that a remote adversary could cause such message loss, for example by causing congestion on targeted links.
Albeit this model is stronger than those of the other bounds in terms of behavior (active), it can as well be considered weaker in terms of the needed eavesdropping capabilities.

\subsubsection*{Note regarding the system models}
Recall that in the service model, the receiver is not an active part of the network but an external entity (like in Tor).
Corrupting a receiver in this case can be achieved in different ways. 
Beyond controlling the receiver herself, for unencrypted traffic it suffices to control only her network links (trivial for the ISP), or the last node on her anonymization path.
Note, that for the \Gelernter\ the traffic to the receiver can even be encrypted as only the fact that those are real messages is important.

We include the adversary model that only eavesdrops on these links  as \emph{service model restricted} to our comparison. It is weaker than those of the \Gelernter\ and \Trilemma. 

\todo[inline]{Alternative:
If we assume, that actively dropping messages on a link is only possible with access to the link, hence includes also passive eavesdropping capabilities on it (footnote:This assumption excludes e.g. attacks that try to introduce congestion on these links remotely.), then we can even argue that the \Ando\ is, in high-level consideration where we ignore the number of links to which the adversary has access to, stronger (we can even do active attacks on sender links, not only passive observation) than our eavesdropping adversary model for the service model case.}


 \subsection{Bounds} \label{sec:mainBounds}
 We explain the minimun cost as inferred by the proofs of the bounds in the following and compare them with each other. Therefore, we unify the notation of the different bounds as follows. 

An ACN has $n$ users. The set $\mathcal{U}$ includes all senders, $\mathcal{U}_H$ the  $h$ honest senders. If the privacy goal challenges the adversary to decide between two suspect senders, we call these ``challenge senders $u_0$ and  $u_1$'' and $u_1$ the ``alternative user to $u_0$''. Further, we refer to the message of this communication as ``critical'' or ``challenge message''. $\lambda$~is the security parameter and $\delta$  the advantage of the  adversary in identifying the real sender.

The \Trilemma\ requires a message to be delivered after at most $l_{max}$ rounds/hops.
We additionally write $l_{exp}$ for the average of the number of hops.

Further, the \Trilemma\ assumes dummy and real messages to be distributed uniformly over several rounds. 
We use $\beta$ to denote the probability that a node sends a dummy message in a given round, $p'$ for the probability of sending a real message. $p=p'+\beta$ is the total probability that a node sends in a round.

$c_p$ ($c_a$) is the number of intermediate nodes the adversary compromised passively (actively). 

Table  \ref{table: notations all } and  \ref*{table: our notations}  of the Appendix show the connection to the original notation and summarize our notation.

\subsubsection{\Gelernter}
Recall the basic idea: If a sender sends less messages than are received, she cannot have been the sender of all these messages. 
We want to prevent the adversary from excluding any sender from the set of suspects that could have sent \emph{all} messages. Thus the number of real messages received can be at most the total number of messages (real and dummy) any one sender has sent. 

More formally, let $Out(r)$ denote\footnote{Compared to \cite{gelernter2013limits}, we omit additional parameters. I.e. $Out(r)$ is short for $Out^\pi_{\sigma,r}$.} the number of messages received by a destination until round $r$ and $L_i(r)$ the number of messages sent by sender $u_i$. The bound is
\begin{equation*}
Out(r) \leq \min\{L_i(r)|u_i \in \mathcal{U}_H\}. \label{equ: Out def.}
\end{equation*}

It follows that the total number of sending events for all senders $Com(r)$ has to be sufficiently high:
\[Com(r)  \geq Out(r) \cdot |\mathcal{U}_H|=Out(r) \cdot h\]

The bound shows a required overhead of at least $h-1$ dummy messages per message that reaches the destination.  In other words, $\frac{h-1}{h}$ of messages are overhead because there are at least $h$ times more sending events than received messages. 
The distribution of overhead during each round is flexible, as long as the sum of the overhead compensates for all delivered messages up to any specific round.

Note that the bound in \cite{gelernter2013limits} is  given in the number of honest senders instead of all senders. As the \Gelernter's adversary model does not include corrupted senders, both numbers are equivalent ($h=n$).

\subsubsection{\Trilemma}
The \Trilemma\ states the trade-offs for two adversaries (non-compromising and compromising), 
and two sending behaviors (synchronized and unsynchronized) and infers areas, i.e. bandwidth-latency combinations, where their privacy goal cannot be achieved.

\paragraph{Non-compromising adversary with synchronized users}
Recall the attack: The adversary knows the two users out of which one has sent the challenge message. 
She also knows the interval, during which the message must have been sent. 
If she does not observe the alternative user sending a message in this interval, she knows the sender with certainty. 
Otherwise, she randomly accuses one of her two suspects. 

Her advantage over guessing randomly equals the probability that the alternative user has not sent a message within the critical interval.
In the synchronized setting, the probability that the alternative user \emph{sent a message} in the critical time interval of $l_{max}$ rounds, is bounded by the sum of the probabilities that she sent a real message\footnote{$\frac{\text{number of rounds (except sending of critical message)}}{\text{number of users (except real sender)}}$}($=\frac{l_{max}-2}{n-1}$) and the probability that she sent a dummy message\footnote{ $\frac{\text{number of dummy messages in the rounds}}{\text{number of users}}$} ($\leq \frac{\beta n (l_{max}-1)}{n-1}$). 
So the bound is simply the probability of the complementary event: 
\begin{align*}
\delta \geq & 1-min\left(1,\frac{(l_{max}-2)+\beta n (l_{max}-1)}{n-1}\right)\\
 \geq & 1-min\left(1,\frac{(l_{max}-1)(1 +\beta n)}{n-1}\right) 
\end{align*}


\paragraph*{Comparison to \Gelernter}
This is intertwined with the considered privacy goals and protocol assumptions.
The \Gelernter\ aims at achieving $S\bar{O}$. To hide the sending frequency for the adversary model\footnote{The \Trilemma\  non-compromising and \Gelernter 's adversary model are the same except for the number of victims.},  everybody else has to send a dummy message, whenever a single user is sending a real message.
The \Trilemma\ aims at $(SM)\bar{L}$, which allows the number of dummy messages to be reduced, as the sending frequencies do not need to be hidden. Sending a real message as alternative sender in the critical time interval is enough to hide the sender-message linking, as desired. This joint sending is reflected in the first part of the sum ($\frac{l_{max}-1}{n-1}$) in the \Trilemma\ for synchronized users, which therefore cannot be found in the \Gelernter. 

When we would require the latency to be minimal for their protocol model ($l_{max}=2$), we force the \Trilemma\ to only consider dummy messages of the current round as suitable cover: As per the protocol assumption in synchronized sending only one real message per round is sent and as the critical interval is just this round, only these dummy messages contribute to the hiding, just as for hiding frequencies in the \Gelernter.  
We also see this reflected in the more precise formula by setting $l_{max}=2$:
\[\delta \geq 1-min\left(1,\frac{\beta n}{n-1}\right)\]
This advantage is $0$ if $\beta=\frac{n-1}{n}=\frac{h-1}{h}$; exactly the overhead required in the \Gelernter.

\paragraph{Non-compromising adversary with unsynchronized users}

We know that an alternative user does not send in a specific round with probability $1-p$.
Additionally, the choice of sending in the $l_{max}-1$ rounds is independent.
We hence can bound the probability that a second observed user does not send by
\[\delta \geq (1-p)^{l_{max}-1}\] (cf. Appendix~\ref{sec:worseTrilemma} for a discussion of the original bound in this case).

\paragraph*{Comparison to \Gelernter}
The \Gelernter\ does not require the overhead to be evenly distributed over the rounds. We however temporarily assume so to allow for a comparison. Thereby the  \Gelernter\  requires the probability of a user sending any (real or dummy) message to be $p=p'+\beta=1$.
This resembles the improved \Trilemma\ bound to minimal latency and the additional effects of a higher latency can be seen in the \Trilemma , but not mapped to the \Gelernter; as in the case with synchronized users.


\paragraph{Non-compromising adversary's area of impossibility}
Based on the above bounds on the advantage, we can infer that for some combinations of latency $l_{max}$ and bandwidth overhead $\beta$ the considered attack has non-negligible advantage, i.e. the privacy goal is broken. 
Such parameter combinations constitute the \emph{area of impossibility}.

If we e.g. assume that the message should be delivered after only one intermediate relay processed it, the adversary wins unless the alternative user sends in the same round (to this relay). Thus unless every user sends in every round ($\beta$ approaches $1$), the goal cannot be achieved for this short latency ($l_{max}=2$).
For the other extreme case of no dummy messages ($\beta=0$), the adversary wins unless the alternative user sends her real message while the challenge message is routed. Thus unless the latency is very high (e.g. $l_{max}=n+1$) and all users send their own message in the meantime with overwhelming probability, the privacy goal cannot be achieved.
 
The \Trilemma\  makes the assumption that $n\approx poly(\lambda)$ and derives the following equations for the synchronized setting. 
All parameter combinations that fulfill them cannot  achieve the privacy goal. 
\[2(l_{max} -1)\beta \leq 1-\frac{1}{poly(\lambda)} , \beta n \geq 1\]

For the unsynchronized setting the equations are equal, except that $\beta$ is replaced with $p$.

\paragraph*{Comparison to  \Gelernter}
The \Gelernter\ and the area of impossibility from the \Trilemma\ can be transformed to the following statements (cf. Appendix \ref{appendix: gelernter and trilemma: equations}): 
\begin{equation*}
\text{\Gelernter: }\beta \geq \frac{Out(r)}{r} \left(1-\frac{1}{poly(\lambda)}\right)\label{eq: summary gelernter}, 
p=1
\end{equation*}
\begin{equation*}
\text{\Trilemma: }\beta  \geq \frac{1}{2(l_{max}-1)}\left(1-\frac{1}{poly(\lambda)}\right), \beta n \geq 1 \label{eq: summary trilemma}
\end{equation*}
$\frac{Out(r)}{r}$ is the average number of messages delivered to the destination in each round.
We hence assume this number not to be much lower than $1$ for most protocols, to retain utility.\footnote{In the Trilemma \cite{das2018anonymity} the number of delivered messages per round $\frac{n}{n+(l_{max}-1)}$ approaches 1 for  high numbers of users, for the protocol used in the \Gelernter\  the number is $1$.} 
Also, Section \ref{sec:trilemmaNoLatency} yields that the Trilemma requires ${l_{max}>1}$.
In consequence, it holds that
$\frac{1}{2(l_{max}-1)}<\frac{1}{2}$. 
This shows that the lower bound on the bandwidth overhead of the \Gelernter\ is higher, reflecting its stricter requirements.

\paragraph{Compromising adversary}
Extending the adversary to compromise up to $c_p\leq n-2$ intermediate nodes facilitates the attack of tracing messages along their anonymization paths, if all nodes on these paths are under adversarial control.
This increases the advantage of the adversary, and the \Trilemma\  is interested in this additional probability for an attack to succeed.
We explain how the probability is bounded in the Appendix~\ref{App:TrilemmaCompromising} and only discuss the \emph{area of impossibility for the compromising adversary} here.

If an adversary passively compromises  $c_p<l_{max}-1$ protocol parties, then the area of impossibility is  \[2(l_{max}-1-c_p)\beta \leq 1-\frac{1}{poly(\lambda)}\text{.}\]
If the number of compromised parties is ${c_p\geq l_{max}-1}$, then anonymity cannot be reached for  \[2(l_{max}-1)\beta \leq 1-\frac{1}{poly(\lambda)} \text{ and } l_{max}\in O(1)\text{.}\]

\paragraph*{Comparison to \Gelernter}
We have already compared the case without compromised protocol parties. Adding them the result cannot directly  be matched to the \Gelernter\ (as it uses no compromised protocol parties). 
We can however again transform the impossibility area for the case of ${c_p<l_{max}-1}$ to
\begin{equation*}
\beta  \geq \frac{1}{2(l_{max}-1-c_p)}\left(1-\frac{1}{poly(\lambda)}\right) \label{eq: summary trilemmaCompromised}
\end{equation*}
Note, that this requires more bandwidth overhead than without compromised intermediate nodes, as expected. It is however still a weaker bandwidth requirement than in the \Gelernter, as $\frac{1}{2(l_{max}-1-c_p)}< \frac{1}{2}$ and the \Gelernter's factor ($\frac{Out(r)}{r}$) is assumed to be close to 1.  For the case of more corrupted parties, interestingly a constant latency is no longer possible as this ensures a non-negligible advantage that cannot be balanced with bandwidth. 

\subsubsection{\Ando} \label{sec:mainBoundAndo}
Recall the  idea for the \Ando: The adversary has two suspects, from which one is sending to the receiver, in which the adversary is interested in. In her attack,  she correlates her dropping of packets sent from the victim with the missing arrival of an expected packet at the receiver. This attack is successful unless one of two cases happens: 1) the communication of the alternative suspect is routed over the victim and thus the adversary wrongly accuses the victim even if the alternative suspect sends to the receiver or 2) the adversary cannot drop all copies of the message sent by the victim and thus wrongly acquits the victim. 

This bound measures \emph{onion cost} of a user as the expected number of packets (own and relayed messages) a user sends. The bound is based on two  key observations: 1) if the victim forwards less packets than users exist (sublinear onion cost in the number of users) there is some user whose packet she does not forward, and 2) if the victim sends only few copies (logarithmic in the security parameter\footnote{They implicitly assume that this is also sublinear in the number of users.}) and the amount of corrupted nodes is high enough,\footnote{They do not mention any requirement on the number of corrupted nodes except that the amount is constant, but if there are less corrupted nodes than copies of the message, she cannot succeed by dropping messages coming directly from the victim.} the adversary can drop all copies with non-negligible probability.
Combining the two observations, the attack leads to a non-negligible advantage. 
 Thus, for their privacy goal, the onion cost per user has to increase faster than $\log \lambda$, i.e. be in $\omega(\log \lambda)$, and therefore the onion cost for the whole network has to be in $\omega(n \log \lambda)$.


\paragraph*{Comparison} 
The authors of the \Ando\ assign the \Trilemma\ an onion cost of $\omega(n)$, while their own result entails $\omega(n \cdot \log \lambda)$ due to the ``stronger''\footnote{It is not stronger in all dimensions as discussed in Section~\ref{sec:adversaryComp}. } active adversary~\cite{ando2019complexity}.

We cannot confirm this. The number of onions that are directly sent, i.e. the onion cost, is the number of onions created by all users multiplied by the rounds they stay in the network. To be able to compare the onion costs, we use $ l_{exp}\leq l_{max}$ as the  average latency of the protocol. The number of onions is the number of real messages plus the number of dummy messages. Both are multiplied with  the number of rounds messages spent in the network ($l_{exp}$). Recall for the \Trilemma: $\beta + p'= p$ messages per user and round are sent. There are $n$ users as well as rounds -- as every user is assumed to send one message, and only one user sends a real message per round. This results in $n^2\cdot p$ messages in total that stay for $l_{exp}$ hops in the network. So the onion cost per user in the Trilemma is $ \frac{n ^2 \cdot p \cdot l_{exp}}{n}  = n \cdot  p  \cdot  l_{exp}\text{,}$
or $n^2 \cdot p \cdot l_{exp}$ for the complete network, which is considerably higher than a bound of $\omega(n)$ or $\omega(n \log \lambda)$.

Considering the impossibility area, the \Trilemma\ states that $(SM)\bar{L}$ is impossible for onion cost of $\frac{n^2(poly(\lambda)-1)}{2poly(\lambda)}\approx n^2$. 

The onion cost for the \emph{\Gelernter} for the whole network can be transformed to $\left(n^2 (\frac{n-1}{n}+\frac{1}{n}) \right) \cdot l_{exp}=n^2 \cdot l_{exp}$. 

Note that also the precise onion cost for \Gelernter\ is higher than for the \Trilemma\ (since $p \leq 1$, $n^2 \cdot l_{exp}\geq n^2  \cdot  p  \cdot  l_{exp}$). Both onion costs are higher than the one for the \Ando, contrary to the claim in \cite{ando2019complexity}.

\paragraph*{Note on the \Ando\ in latency and bandwidth overhead}
On the other hand, we can translate the onion cost of the \Ando\ into a bound on latency and bandwidth overhead under the assumption that every message stays in the network for the allowed latency. The resulting impossibility area confirms the above order in costs:
\[p\cdot n\cdot l_{max} > \log \lambda \iff  p >\frac{ \log \lambda}{poly \lambda \cdot l_{max}}\]

\subsubsection{Intermediate Summary on the Overhead Comparison}
All papers discuss the influence of bandwidth overhead on anonymity. The latency overhead is explicitly considered in the \Trilemma\  and implicitly in the \Ando. 
To permit a comparison between the bounds, we transformed all bounds to account for the differing models and assumptions. In result, the overhead required by the \Gelernter\ and \Hevia\ is the highest, by the \Trilemma\ the second highest and the \Ando, albeit based on an active adversary, introduces the lowest overhead. 

 \subsection{Summary}

\begin{table*}[ht]
	\caption{Final Bounds Summary}
	\center
	\resizebox{0.95\textwidth}{!}{%
		\begin{tabular}{ c | p{1.2cm} | p{1.5cm} | p{4cm} |  p{4.5cm} |  p{4.7cm} }
			\hline
			Paper & Notion & Adversary & Protocol Assumptions & Attack &Bound (comparable case, formal)  \\
			\hline\hline			
			\cite{gelernter2013limits} & $S\bar{O}$ & eavesdrop + receiver & no restriction & count messages sent  from victim; if more received they are not all from the victim & $\beta \geq \frac{Out(r)}{r} \left(1-\frac{1}{poly(\lambda)}\right)$ , $p=1$\\
			\hline
			\cite{hevia2008indistinguishability} & $S\bar{O}$ & eavesdrop & guaranteed delivery for up to $\mu_{max}$ messages & count  messages sent from victim; if more received they are not all from the victim & $\beta \geq \frac{Out(r)}{r} \left(1-\frac{1}{poly(\lambda)}\right)$, $p=1$\\
			\hline
			\cite{das2018anonymity} & $(SM)\bar{L}$ & eavesdrop + receiver + relay &required message delivery after $l_{max}$ rounds; onion routing, mix nets, not applicable for DC-Nets & exclude senders that did not send in the time where the critical message was sent, if all relays corrupt: trace message&$\beta\! \geq \frac{1}{2(l_{max}-1)}\left(1-\frac{1}{poly(\lambda)}\right), {\beta n \geq 1} $\\
			\hline
			\cite{ando2019complexity} & $(SR)\bar{L}$ & active + receiver &delivery guaranteed (unless aborted), only onion routing, mixing & drop all messages send from the victim observe missing of expected message at receiver& $p >\frac{ \log \lambda}{poly \lambda \cdot l_{max}}$ \\
	\end{tabular}}
	\label{tab:differences}
\end{table*}

Interestingly, even though all bounds come to similar conclusions, their privacy goal, attacker model, protocol assumptions and also postulated cost differ considerably (cf. Table~\ref{tab:differences}). 

For the \Hevia\ and the \Gelernter\ we realized that cost and privacy goal are equal, and that the attacker models differ only slightly. The differences are easily explained by deviating protocol assumptions. 
Resulting, the \Hevia\ and \Gelernter\ have the highest cost and discuss the strongest privacy goal, albeit in face of a comparably weak adversary model, and without (\Gelernter) or with minor (\Hevia) restrictions on the protocol types. 
Thus, protecting $S\bar{O}$, which explicitly hides which sender sends how often, against an adversary that both observes the first link of the sender (e.g. her ISP) and the corresponding receiver (or has external knowledge about the number of received real messages) is indeed only possible with high bandwidth overhead. 

The papers introducing the other two bounds state stronger, yet analyze lower privacy goals, and postulate lower cost.
The \Trilemma\ aims at unlinking the sender from her message ($(SM)\bar{L}$), while the \Ando\ aims at unlinking pairs of senders and their receivers ($(SR)\bar{L}$). 
Note that while for $S\bar{O}$  every sender sends a dummy message per real message  to assure that real sending frequencies are hidden, for $(SM)\bar{L}$  (or $(SR)\bar{L}$) the bandwidth can be lower as it allows to learn that someone is a more active sender, as long as one cannot link a certain message (or receiver) to her. 
On the other hand, compared to the \Hevia\ and \Gelernter, the adversary model in the \Trilemma\ is slightly stronger. Interestingly, reducing the privacy notion, but using a slightly stronger adversary model for the \Trilemma\ (and an incompatible adversary model for the \Ando), allows the bounds on the overhead to drop considerably. So, in this case the change in the adversary model cannot outbalance the change in the privacy goal.

The cost for unlinking sender and message in the \Trilemma\ is higher than unlinking sender and receiver in the \Ando\, although the latter assumes an active adversary.
The reasons are that the \Trilemma\ is tailored to this special case and that timing observations are exploited\footnote{ Even though the \Ando\  argues that dropping is the most important attack vector as a timeout causes delayed messages to be dropped and modified onions cannot be peeled by the next relay, we suspect that timings cannot be handled that easily with timeouts. Even smaller delays that do not cause a timeout might be recognizable by the adversary or otherwise we need to timeout so early that we expect it to threaten availability.} in the \Trilemma. Further, the adversary in the \Trilemma\ is not strictly weaker than the one for the \Ando. They indeed are incompatible, as the latter is stronger with respect to its behavior being active, whereas the former has a larger area of control, as it can compromise more and different parts of the network.

 \section{Implications}\label{sec:implications}

We extend the idea from \cite{das2018anonymity} to contextualize our results with existing \acp{ACN}.
The comparison to actual ACN protocols of course has to be taken with grains of salt: Exceeding the theoretical bounds in overhead indicates that an ACN {\em may}, but not that it actually {\em does} achieve the corresponding privacy notion.
We discuss system classes, loosely ordering them by the extent to which they can meet the different bounds.


\subsection{Discussion of Networks} 
\begin{figure}[]
	\center
	\includegraphics[width=0.38\textwidth]{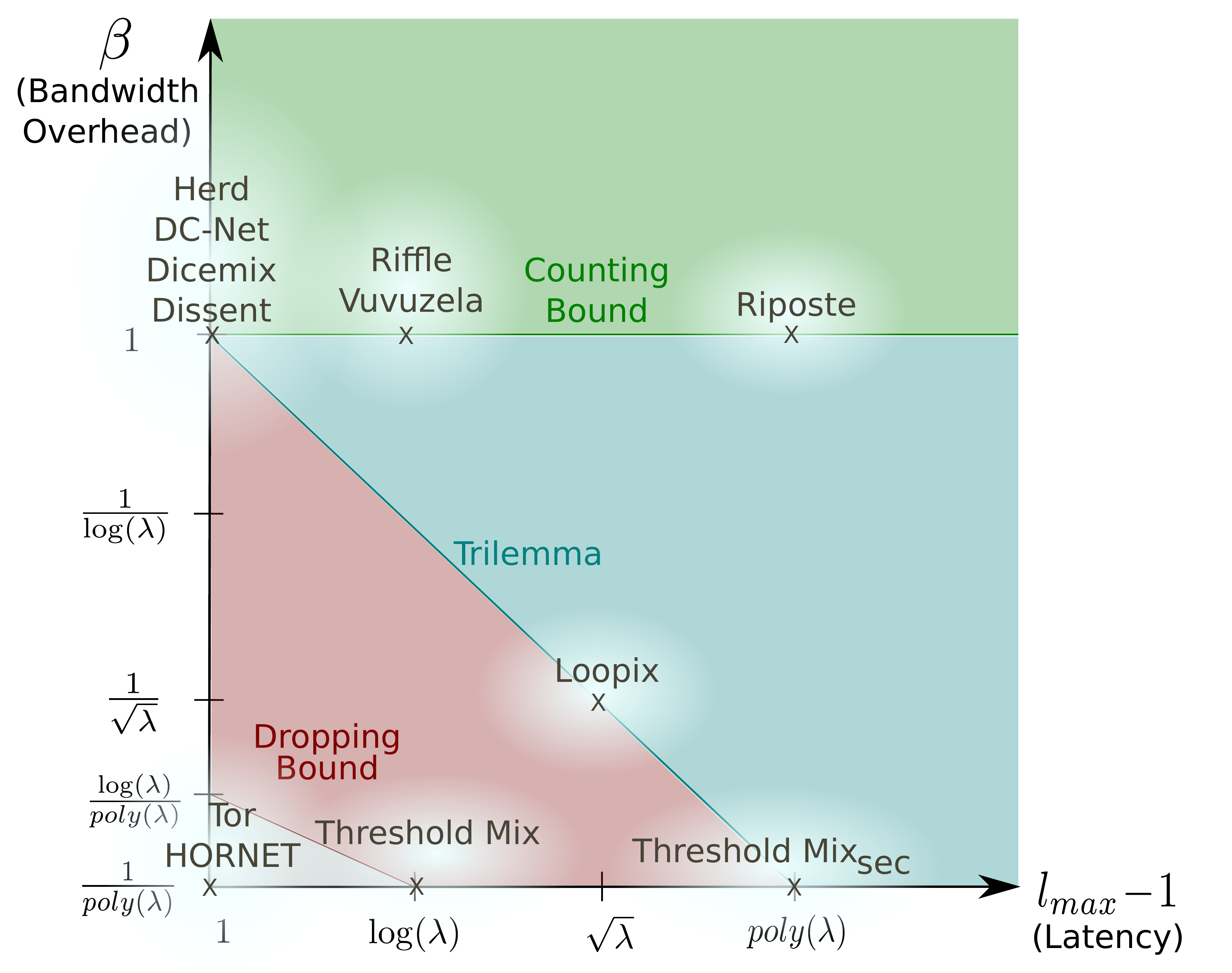}
	\caption{Comparison of bounds under the special set of assumptions of the \Trilemma\ \cite{das2018anonymity} (see Table~\ref{tab:sending} for the \Gelernter\ in the general case).
	As \cite{das2018anonymity} we assume that $\beta \approx p$ to summarize both user settings: The \Gelernter\ requires the highest overhead but is independent of the latency. The \Trilemma\ shows a trade-off between latency and bandwidth, it is higher than the \Ando. }
	\label{fig:triangle}
\end{figure}

Figure \ref{fig:triangle} illustrates the trade-off vs. the cost of different existing ACNs.
We facilitate this comparison  by restricting ourselves to a specific scenario:
A single real message is assumed to be sent during each round.
The abscissa denotes the latency of messages, and the ordinate the bandwidth overhead, as part of the probability that a node sends a  message during a round.

We also discuss the more general cases, which are especially interesting to assess the systems according to the \Gelernter\ -- and we give an overview  on the assumptions of the different protocols  necessary to assess this in Table \ref{tab:sending}.

\begin{table}[]
	\caption{Comparison of assumed sending behavior }
	\resizebox{0.47\textwidth}{!}{
		\begin{tabular}{ c  c   c   c  }
			\hline
			\textbf{Protocols} & \textbf{Dummy traffic} & \textbf{Communications} & \textbf{Counting-  }  \\
			\textbf{} & \textbf{ per round \& user} & \textbf{ per round } & \textbf{Bound }  \\
			\hline
			Herd& 1 (some) & n (or more) & X \\ 
			\hline
			DC-Net, Dissent & 1& 1& \checkmark\\ \hline
			Dicemix & n &n & \checkmark \\ \hline
			Vuvuzela & 1 & n&  X \\ \hline
			Riffle, Riposte & 0 (1) & n & X \\ \hline
	\end{tabular}}
	\label{tab:sending}
\end{table}

\subsubsection{Tor \cite{dingledine04tor}, HORNET \cite{chen_hornet:_2015}}
This first class of low overhead onion routing systems sends messages over a path of relays and does not employ additional dummy traffic. The number of hops is fixed, and thus they expose constant latency. 

These systems fall short of any bound. All explained attacks indeed are successful: It is simple to link sent onions because of their timing  (\Trilemma), to count the number of messages a sender sent  as sending can be observed (\Gelernter). 
Knowing that a certain receiver expects another packet (e.g., because the use case postulates a message stream), dropping it right at the sender can be recognized at the receiver (\Ando).
This in itself is not new, and corresponding attacks have been suggested~\cite{murdoch2005low} or are at least conceivable for Tor.

\subsubsection{Threshold-Mix \cite{serjantov2002trickle}} This class of mixes collects $t$ messages before relaying them further.
It does not employ dummy traffic. 
Thus, each sending event transmits a real message, and $S\bar{O}$ cannot be achieved according to the \Gelernter. 
Interestingly, the approach can however fulfill the  two other bounds: If  each user sends one message and each mix waits for all of them, and if further all mixes are used (as assumed by both bounds if a high latency is allowed), the attacks fail. 
Dropping a message yields no message to be delivered, and hence the privacy is kept (although availability is jeopardized).\footnote{Threshold mixes do not employ any additional technique to protect against the attack of the \Ando.}
As long as we assume that one of the mixes is honest, linking the incoming and outgoing packets fails at this point and also timing does not provide any help as the first mix already waits for all messages. We do however agree with \cite{das2018anonymity} that the \Trilemma\ and \Ando\ cannot be met for convenient thresholds and numbers of mixes.

\subsubsection{Herd \cite{le2015herd}, DC-Net \cite{chaum1988dining}, Dicemix \cite{ruffing2017p2p}, Dissent \cite{wolinsky2012dissent}}
This class of systems employs dummy traffic but has low latency. 
Herd uses multiple relays just like Tor and HORNET, but adds dummy traffic. DC-Net, Dicemix and Dissent in contrast follow the idea of superposed sending. They generate the original message as a combination of both; a real message from one and dummy messages from all other users. 

Only the \Gelernter\ is applicable to these superposed sending based systems, as both the \Ando\ and the \Trilemma\ are based on the mixing model, which includes Herd but not systems based on superposition.


The systems indeed meet the overhead requirement of the \Gelernter.
Without a collision avoiding scheme (cf. \ref{sec:backDC}) DC-Nets still cannot achieve the notion $S\bar{O}$ (cf. \cite{gelernter2013limits}).
Dicemix and Dissent specify scheduling for transmission slots by combining one message of each user in every round.
Mounting the attack from the \Gelernter, the receiver hence does learn that all messages of a single round are from different senders, and only messages distributed over multiple rounds can be from the same sender.
She succeeds and the notion $S\bar{O}$ cannot be achieved in consequence.


The situation for Herd is a bit more complex, than the representation in Figure \ref{fig:triangle} suggests.
The graph assumes only a single communication per round, and for this special case Herd meets all bounds as it employs enough overhead.
However, Herd aims at a VoIP scenario, which indicates that the more general case of users participating with several communications in the same round seems more applicable.
The \Gelernter\ is no longer met in this case:
The users in Herd generate a predefined amount of traffic, which is supposed to at least resemble the traffic caused by a small number of  VoIP connections (e.g. one). 
This does not outweigh the total number of real messages sent during a round, and the \Gelernter\ is violated.
Herd in consequence leaks some information about the sender behavior and $S\bar{O}$ cannot be achieved.


\subsubsection{Loopix \cite{piotrowska2017loopix}}
Loopix is another mix network that adds more mixes and dummy traffic. 
It allows to adjust both the number of used mixes and the dummy traffic via parameters. 
Sender traffic is generated according to an exponential distribution. 
Like \cite{das2018anonymity} we assume $\sqrt{\lambda}$ mixes per path  and dummy traffic with probability $\frac{1}{\lambda}$, although we  stress that other parameter choices are not excluded by the paper.

Loopix in this setting satisfies nearly all bounds. 
Only from the \Gelernter, we can conclude that it cannot achieve $S\bar{O}$. 
We expect that also practically the two scenarios of either one user sending many messages, or the same number of messages being sent by multiple users can be distinguished:
The messages in the first case arrive much slower at the receiver.
However, aiming at $S\overline{O}$ may be too strong for many use cases and a weaker privacy notion targeted. 
Loopix does employ protection measures against the other attacks.
Confirming their effectiveness is beyond the scope of this paper, and we leave it for future work.

\subsubsection{Riposte \cite{corrigan2015riposte}}
Riposte uses a reversed PIR to implement an anonymous broadcast. 
Each client sends a message to the PIR servers in the epoch during which she participates. 
Riposte does not apply the concept of dummy messages\footnote{Only for receiving an empty message is used, as messages in the postboxes of the clients are swapped.}. 
The set of senders is published at the end of each round.

Riposte does not lend itself to analysis with the model of the \Trilemma, as the latter assumes only a single sender to send a real message per round, but Riposte requires several parallel communications to achieve any anonymity.
In Figure~\ref{fig:triangle} we still follow \cite{das2018anonymity} and choose the probability for every sender to send in each epoch to be one.


Categorizing Riposte with a sending probability of 1 is misleading for the general case as not only one, but multiple messages are sent per epoch.
Similar to Herd, the bandwidth overhead is again too small to withstand the requirements of the \Gelernter. 
We can confirm this with a practical attack: By observing the number of write requests to the servers (i.e. send events), an adversary can directly count the number of sent messages, as no dummy traffic is applied. 
Riposte clusters sending events, so they are not spread over several rounds, and they are only hidden among each other.
Further, although the latency is sufficient to fulfill the \Ando, the dropping attack still works: Dropping all parts of the write request of one user will not lead Riposte to stop, but instead to publish all except this user's message.

\subsubsection{Riffle \cite{kwon2016riffle}, Vuvuzela \cite{van2015vuvuzela}}
Vuvuzela and Riffle are mix networks that require all messages to go trough all mixes. 
Alike \cite{das2018anonymity}, we assume a logarithmic number of mixes. 
Further, Vuvuzela ensures a constant traffic rate by employing dummy traffic. Riffle assumes all clients to always have a message to send (``each client onion encrypts a message''). So, in both protocols each client sends in every round.  Riffle additionally employs PIR to deliver the messages after they went through a verifiable shuffling mix net. 

They intuitively seem to satisfy all bounds and could possibly achieve all notions. 
Similar to Herd however, multiple users can (Vuvuzela) or  have to (Riffle) send every round and we can infer that all messages of one round have been sent by different senders. 
Thus the \Gelernter\ is only fulfilled for the special case that only one user sends per round. This case might happen, but is not enforced in Vuvuzela, and contradicts the assumption of Riffle that each client sends a message. 




\subsection{Summary}
The bounds show limitations of existing ACNs, as they cannot achieve certain privacy notions.
We managed to underline this situations with real-world attacks on the systems.
We also conclude that nearly no system achieves $S\bar{O}$ nor reaches the \Gelernter\ under the given assumptions.
It turns out that the assumption of the number of real messages sent per round is important. Not only to assess specific bounds and check for their applicability in the first place, but also to put the bandwidth overhead into perspective.

It remains to state that there are cases where we suspect that the protocols do not achieve certain privacy goals even though they reach the corresponding bounds.


\section{A Practical Viewpoint: Explaining Limitations}\label{sec:practical}
Arriving at this bleak outlook, we want to put the bound into perspective.

\subsection{Strong Privacy Goal Formalizations}
\subsubsection{The Notion $S\overline{O}$ of the \Gelernter}
$S\bar{O}$ is a strong notion\footnote{Note, that also much stronger notions, which require for instance membership concealment, hiding the fact if a user participates in the system at all, are discussed in literature.}, which even hides the number of active senders. 
While there are use cases for this notion \cite{kuhn2018privacy}, for many proposed protocols it might be too strong. 
Some protocols (cf.  Section \ref{sec:implications}) aim however to protect against a similar, but weaker notion\footnote{For a formal definition of this weaker notion see $S\overline{O}_{n_{\max}}$ in Appendix \ref{RelaxedSO} and for further useful, weaker notions see \cite{kuhn2018privacy}.}: 
They ensure that any user sends a fixed, small number of communications (real or dummy) every round. 
Thereby, they allow the adversary to learn that no user has sent more than this number of real messages, which implicitly leaks a lower bound on how many senders have been active during a given round.

\subsubsection{Game-Based Notions for Bounds}

Everything that \emph{could} leak in the protocol by definition of the game-based notion is assumed to \emph{be} leaked during the analysis. 
This is useful for worst case analyses. 
For bounds, however,  the adversary knows, per game definition, everything that happens as long as it is not explicitly defined to be protected. She does not even have to be able to observe any of this in reality.

Consider the \Hevia: The adversary knows how many real messages are received, without controlling the receiver, just by the definition of the notion. 
Further, the attack in the \Ando\ requires her to realize that a packet is missing.
In the game-based notion, this is trivial: the adversary knows how many messages each receiver expects, by the definition of the notion.
In reality, this limits the applicability to use cases with predictable receiving behavior (like streaming or triggering the reaction with rumor spreading).

Future work on bounds should therefore argue the practicability of the underlying attack and assumptions.
For more realistic analyses communications unknown to the adversary and beyond her control could be included\footnote{This extension is easily achieved by adding adversary classes~\cite{backes17anoa, kuhn2018privacy} }.

\subsection{Maximal Anonymity Sets}
All bounds require the anonymity set to include all users and that even the considered attack cannot exclude a single user from it. For many real use cases, however, significantly smaller anonymity sets after an attack may be sufficient. 
For example, building the anonymity set only from the users concurrently online (or sending) might be acceptable for the use case as long as at any point \emph{enough} users are online (or sending). Determining such suitable smaller anonymity sets will be a challenge for future work.

\subsection{Bandwidth Cost Models}

Different concepts are summarized under the term ``bandwidth overhead''. 
For the \Trilemma\ bandwidth overhead naturally occurs from dummy messages, while for the \Ando\ redundant copies of the real messages are needed.  Further, also for dummy messages end-to-end dummy traffic, starting and ending at users, and link dummy traffic, which is just applied to obscure the traffic on one hop, exist. 

Interestingly, the overhead in the \Gelernter\ and  \Trilemma\ measures only in the sender-generated dummy messages.  
In practice, however, end-to-end dummy traffic puts more load on the network than link dummy traffic at the sender's first link.  In the \Trilemma, for example, longer lasting dummy traffic would only be necessary if corrupted relays are introduced into the model.
Contrary to the cost definition of the \Gelernter\ and \Trilemma, the \Ando's can reflect a difference between end-to-end dummy traffic and dummy traffic on the first link. We thus prefer this cost metric for future work.
 
 \subsection{Assumptions}
 Relaxed assumptions are desirable  for future work on  bounds to improve their applicability.
 In terms of sending behavior, having more than a single user send a real message per round, contrary to the \Trilemma 's assumption, suits reality better and naturally benefits the privacy. Further, latency requirements, like in the \Trilemma , may also be more relaxed in many practical use cases.

\section{Conclusion and Future Work}\label{sec:conclusion}
We have systematized different analyses that prove lower bounds on the overhead that is necessary to achieve certain privacy goals of anonymous communication.
Analyzing their assumptions we have shown that their underlying attacks suffice to break much weaker than the targeted privacy notions, and hence tightened the given bounds.

Presenting the complete landscape of existing bounds, we found that in terms of the adversary all state global capabilities, while the actual attacks only require local influence or observations close to both endpoints of the communication. Only the \Ando\ uses active capabilities, while the others are strictly passive and, except for corrupted intermediate nodes, quite similar.
 All primarily targeted goals protect the sender, but in different ways. While one class (the \Hevia\ and \Gelernter) analyzed the strongest notion that only focuses on the sender, another (the \Trilemma\ and \Hevia) actually investigated two of the weakest goals imaginable. 
 The first class needs no additional restrictions, while the second is only applicable for a subclass of all \acp{ACN} and the \Trilemma\ even makes further assumptions on the sending behavior. 
The resulting overhead requirements for the first class are independent of the acceptable latency. The second class on the other hand shows a trade-off between latency and bandwidth. 
Stricter requirements for the privacy protection lead to higher overhead bounds, even though the adversary model was slightly weaker. Also assumptions on the sending behavior and exploitation of time in the attack resulted in higher required overhead, even compared to another attack exploiting active capabilities.

A critical assessment of the assumptions of the corresponding papers revealed limitations from a practitioner's perspective.
They commonly require the protocol to create a single anonymity set containing all users, even when attacked.
Some assume that only a single real message is sent per round, and corresponding attacks seem harder in reality.
Some proposed cost metrics neglect how often messages are forwarded on the network, and hence do not favor more efficient link-based over end-to-end dummy traffic.

We firmly believe in the utility of treating anonymous communication formally, and proving corresponding efficiency bounds. Our comparison allows practitioners to take the existing knowledge of the bounds for the specific cases in which they apply into account.  
For future work on bounds, we suggest to help identifying the weakest possible assumptions, by stating them more expressly and explaining them from a practical viewpoint, and to improve utility we suggest to consider the discussed practical limitations, by leveraging more realistic cost models, relaxed privacy goals and more realistic assumptions about sending behavior and prior knowledge of the adversary.

\section*{Acknowledgement}
This work in part was funded by the German Research Foundation (DFG, Deutsche Forschungsgemeinschaft) as part of Germany’s Excellence Strategy – EXC 2050/1 – Project ID 390696704 – Cluster of Excellence “Centre for Tactile Internet with Human-in-the-Loop” (CeTI) of Technische Universität Dresden and the Helmholtz Association (HGF) through the Competence Center for Applied Security
Technology (KASTEL).

\bibliographystyle{abbrv}
\bibliography{articles}

\appendix

\section{Formal Privacy Definitions}\label{app:privacyNotions}
\subsection{Basic Goal Formalization from~\cite{kuhn2018privacy}} \label{app:additional}
In game-based security definitions,  the game adversary chooses two \emph{batches}, i.e. sets of communications that start in a random order (or simultaneously). 
The challenger picks one of them at random, simulates it, and provides observations to the adversary, according to the protocol and adversary model. 
This process can be repeated.
The game adversary finally has to guess, which batches were chosen, based on her observations. 
If the adversary learns any information that was required to be hidden, she will be able to distinguish the scenarios\footnote{We use "scenario" to refer the challenge scenarios as in the \Gelernter~\cite{gelernter2013limits} and address different properties of users  and attackers as "settings", contrary to calling them ``scenarios'' like in the \Trilemma~\cite{das2018anonymity}. } and to guess correctly. Thus, her \emph{advantage} in the game, i.e. the improvement of her success probability over random guessing, is non-negligible. The required privacy is hence not achieved, and the notion is considered to be broken.

We can vary this game based on how the adversary is allowed to define the scenarios, i.e. choose the batches.
We restrict her choice, and she has to provide two scenarios that are equal in the information that is not confidential, and hence allowed to leak.
They may only differ in information that is required to be hidden, so her guess depends on her ability to learn protected information.

Creating a hierarchy, the adversary is not restricted in the definition of the strongest goal,
\emph{Communication Unobservability} ($C\bar{O}$): Anything, even how many communications are happening, has to be hidden. 

While $C\bar{O}$ protects both senders and receivers, \emph{Receiver Unobservability} ($R\bar{O}$) is the strongest goal that only protects the receivers. Everything about the senders and their messages can be learned by the adversary and hence is required to be equal in both scenarios.

Similarly, \emph{Sender Unobservability} ($S\bar{O}$)  is the strongest goal that only protects the senders. Everything about the receivers and their messages can be learned by the adversary and hence is required to be equal in both scenarios. 

Relaxing $S\bar{O}$, the weaker goal \emph{Extended Sender-Message Unlinkability}\footnote{$SM\bar{L}$ is called Sender-Message Unlinkability in \cite{kuhn2018privacy}.}   ($SM\bar{L}$) allows the adversary to learn how many messages each sender sends in addition to everything about the receivers and messages. So the (real message) sending frequency of a sender has to be the same as of this sender in the other scenario.

\emph{Sender-Message Unlinkability}\footnote{$(SM)\bar{L}$ is called Pair-Sender-Message Unlinkability, $(SR)\bar{L}$:  Pair-Sender-Receiver Unlinkability in \cite{kuhn2018privacy}.} ($(SM)\bar{L}$) is one of the weakest sender goals. 
It protects only the fact that a message and its sender belong together. 
The second scenario is entirely equal to the first in this case, except that the senders of two communications with the same receiver are exchanged. 
Similarly, \emph{Sender-Receiver Unlinkability}\footnotemark[2]\todo{check footnote} ($(SR)\bar{L}$) only protects the fact that  a sender and receiver communicate together. So, the senders of two communications with the same message are switched in the scenarios.

For the  goal  \emph{Message Unobservability with Message Unlinkability} ($M\bar{O}[M\bar{L}]$), we require that how many messages each sender sends and each receiver receives is equal and does not have to be hidden.

\subsection{Definitions} \label{RelaxedSO}
We explain slightly simplified versions of the definitions from~\cite{kuhn2018privacy} and add a relaxation $S\overline{O}_{n_{\max}}$.
$r=(u,u',m,aux)$ denotes a communication, i.e. message $m$ is send from $u$ to $u'$ with auxiliary information $aux$. Multiple communications are grouped into batches $\underline{r}$. Notions are defined by stating which batches $\underline{r}_0, \underline{r}_1$ have to be indistinguishable, i.e. are allowed to be chosen by the adversary for the challenge.

\begin{figure}[tbh]
  \centering
  \includegraphics[width=0.2\textwidth]{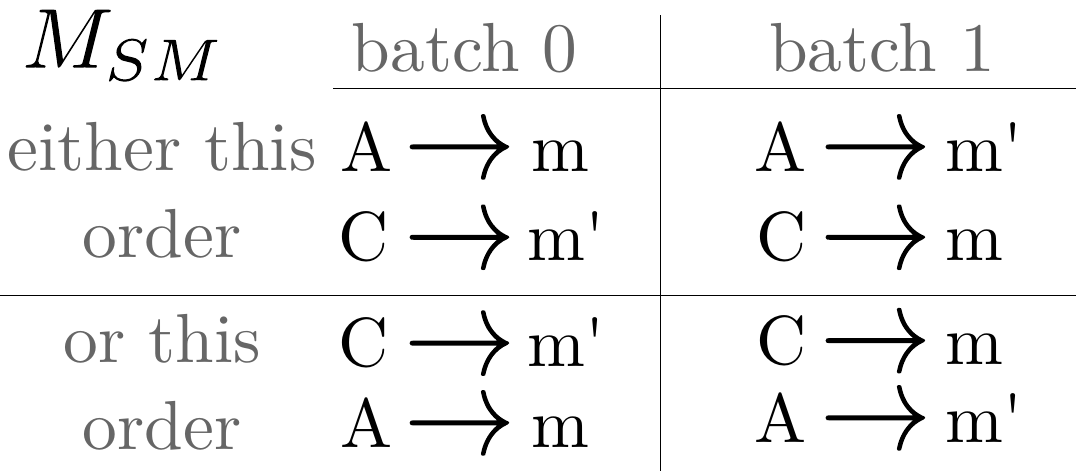}
  \caption{Batches in $M_{SM}$ illustrated as in \cite{2019arXiv191013772K}}
  \label{MSM}
\end{figure}

To define the notions we use  $\underline{r}_0, \underline{r}_1$ for the batches in question, which for $b\in \{0,1\}$ contain communications \[r_{b_j}\in \{(u_{b_j},u'_{b_j},m_{b_j},aux_{b_j}), \Diamond\}\] where $\Diamond$ denotes that no communication is happening.\todo{``]'' in item}
\begin{itemize}[leftmargin=1.5cm]
\item[$C\overline{O}$:] All batches $\underline{r}_0, \underline{r}_1$ are valid.
\item[$M\overline{O}[M\overline{L}$]:] Let $Q_b:=\{(u,n)\mid$ $u$ sends $n$ messages in $\underline{r}_b\}$ denote how many messages each sender sends and $Q'_b$ how many each receiver receives. The batches $\underline{r}_0, \underline{r}_1$ are valid iff $Q_0=Q_1$, $Q'_0=Q'_1$, $\Diamond \not \in \underline{r}_0$ and $\Diamond \not \in \underline{r}_1$.
\item[$R\overline{O}$:]The batches are valid iff for all $j$: \\$r_{1_j}=(\mathbf{u_{0_j}},u'_{1_j},m_{0_j},aux_{0_j})$. 
\item[$S\overline{O}$:]The batches are valid iff for all $j$: \\$r_{1_j}=(\mathbf{u_{1_j}},u'_{0_j},m_{0_j},aux_{0_j})$. 
\item[$S\overline{O}_{n_{\max}}$:]The batches are valid iff for all $j$: \\$r_{1_j}=(\mathbf{u_{1_j}},u'_{0_j},m_{0_j},aux_{0_j})$ and for $b\in \{0,1\}$ for all $(u,n) \in Q_b$: $n\leq n_{max}$. 
\item[$SM\overline{L}$:] The batches are valid iff   for all $j$: \\$r_{1_j}=(\mathbf{u_{1_j}},u'_{0_j},m_{0_j},aux_{0_j})$ and $Q_0=Q_1$.
\item[$(SM)\overline{L}$:] Let $M_{SM}$ specify that only the senders of two messages are swapped in the two batches (see Fig.~\ref{MSM}). Batches are valid iff for all $j$: \\$r_{1_j}=(\mathbf{u_{1_j}},u'_{0_j},\mathbf{m_{1_j}},aux_{0_j})$ and $M_{SM}$ is true. 
\item[$(SR)\overline{L}$:] similar to $(SM)\overline{L}$.
\end{itemize}

\paragraph*{Single Setting}
For reasons of compatibility with the analyzed papers, we extend \cite{kuhn2018privacy} by introducing 
an $X_1$ for each notion $X$. It expresses that every sender sends exactly once in each batch for a sender notion ($S\bar{O}$, $(SM)\bar{L}$), each receiver receives exactly once for a receiver notion ($R\bar{O}$),  and  each sender and receiver send/receive exactly once in each batch for an impartial notion ($C\bar{O}$, $(SR)\bar{L}$).

Formally, the extension$X_1$  is defined to any notion $X$ as for any
\begin{itemize}[leftmargin=2.7cm]
\item[sender notion] for all $b\in\{0,1\}, (u,q)\in Q_b:q=1$, i.e. all users send exactly once in the batch.
\item[receiver notion] for all $b\in\{0,1\}, (u,q)\in Q'_b:q=1$, i.e. all users receive exactly once in the batch.
\item[impartial notion] for all $b\in \{0,1\}, (u,q)\in Q_b\cup Q'_b:q=1$, i.e. all users send and receive exactly once in the batch. 
\end{itemize}

Note that this only expresses weaker privacy goals and in terms of bounds this means, the bound for the goal without this extension is also valid, but slightly less precise.

\subsection{Introducing additional restrictions}
The framework introduces additional concepts: It allows to use the observations of \emph{multiple batches}. Thus,  the game adversary can decide on the next batch after observing the output to the current batch. 
A batch is thereby understood as a sequence of communications, but the semantics of a batch are not defined further in~\cite{kuhn2018privacy}. For this work, we understand a batch as communications that start in an unpredictable order, at least for the adversary. The easiest way is to think of them being initiated simultaneously or in a random order. Formally, this requires using a random permutation over all communications of the batch.

The second concept is called \emph{number of challenges}. It measures how different the two sequences of batches are in terms of how often e.g. in $S\bar{O}$ the senders differ or how many sender pairs have been switched in $(SM)\bar{L}$. Further, the number of challenge rows counts the number of differing communications between the scenarios.

A third additional concept are \emph{corruption restrictions}. Here we are interested in $X_{c^e}$, that specifies that the messages that are sent and received by corrupted users have to be equal in both scenarios as otherwise the adversary could trivially break the notion by observing the behavior at corrupted users. Formally, let $\hat{\mathcal{U}}$ denote the set of corrupted users: for all $\hat{u} \in \hat{\mathcal{U}}$ and all communications $r_{b_j}$ with $\hat{u}$ as sender or receiver:  $m_{1_j}=m_{0_j}$.


\subsection{Comparing advantage definitions}
The advantage definitions of the papers are equivalent. We show equivalence mostly with simple transformations. Only the \Ando\ represents a slight exception, but also its chosen total variation distance can be shown equivalent using known results~\cite{cuff2016differential}.

\subsubsection*{Detailed Comparison}
We use $Pr[g=\mathcal{A}|\mathcal{C}(b)]$ short for the probability that the attacker $\mathcal{A}$ guesses $g$ when the challenger $\mathcal{C}$ picked random bit $b$.

 \paragraph{\Gelernter}
This definition requires the probability that any adversary algorithm $\mathcal{A}$ correctly guesses that bit $b=1$ ($Pr[1=\mathcal{A}|\mathcal{C}(1)]$), is only negligibly bigger than the same algorithm guessing $b=1$ incorrectly ($Pr[1=\mathcal{A}|\mathcal{C}(0)]$). So, the adversary has only a negligible advantage $\delta$ in winning the game.
\[Pr[1=\mathcal{A}|\mathcal{C}(1)] - Pr[1=\mathcal{A}|\mathcal{C}(0)] \leq \delta \]

\paragraph{\Hevia} The definition of the adversary's attack advantage $\delta$ in the \Hevia\ can be shown to be equivalent, under the assumption that the adversary always guesses something, with simple transformations:
%
\begin{align}
\delta_{\Gelernter}&=\Pr(1|1)-\Pr(1|0)\nonumber\\
&=\Pr(1|1)-(1 -\Pr(0|0))\nonumber\\
&=2 \cdot \left(0.5 \cdot \Pr(0|0) + 0.5 \cdot \Pr(1|1)\right) -1\nonumber\\
&= 2 \cdot \Pr(b|b) -1\nonumber= \delta_{\Hevia} 
\end{align}

\paragraph{\Trilemma}
The \Trilemma\ uses a definition similar to the \Gelernter's as follows:
\[Pr[0=\mathcal{A}|\mathcal{C}(0)]\leq Pr[0=\mathcal{A}|\mathcal{C}(1)]+\delta \]
Note that in comparison to the \Gelernter's here only the bits are changed. 
Technically, the advantage definition has to be fulfilled for any PPT adversary. Hence, if there is an adversary violating the definition of the \Gelernter, we can simply swap its chosen scenarios and invert the output bit and we have an adversary violating the definition of the \Trilemma\ and similarly for the other way round.

\paragraph{\Ando}
This anonymity defines two scenarios to be indistinguishable iff the statistical distance between the observation of the adversary is negligible in the security parameter. As measure for the statistical distance the total variation distance ($\Delta_{TV}(\cdot, \cdot)$) is used. 
\[\Delta_{TV}(V_\mathcal{A}|\mathcal{C}(0), V_\mathcal{A}|\mathcal{C}(1)) \leq \delta\]
From \cite{cuff2016differential} (Equation 8) we know that this total variance based definition and the differential privacy based $(0,\delta)$-closeness definition of \cite{cuff2016differential} are interchangeable.  Further, the $(0,\delta)$-closeness definition is defined as the outputs of the mechanism, i.e. the input to the game adversary, being indistinguishable, just as the probabilities in the definition of the \Gelernter\ and \Trilemma:
$ V_\mathcal{A}|\mathcal{C}(0) \leq V_\mathcal{A}|\mathcal{C}(1) + \delta$.

 Thus, the only remaining difference between the definitions is that the \Ando's is talking about the probability distributions in the views of the game adversary, while the \Gelernter's is talking about all possible game adversary algorithms. However, if the difference of the probability distributions in the views is negligible, so is the chance of any adversary to distinguish them. Also, if there is an adversary that can distinguish the scenarios, then the probability distributions in the views have to be non-negligibly different.


\subsection{Comparing additional restrictions}\label{app:additionalRestrictions}

\paragraph{Corrupted users}
The \Ando\ introduces the additional restriction that corrupted users send and receive the same messages in both scenarios, as the scenarios would otherwise be trivially distinguishable. Formally, this matches the corruption restriction  and leads to $(SR)\bar{L}_{1,c^e}$. For our comparison we can however simply add this restriction to all notions. For the other bounds it does not change anything as it is already always fulfilled: all senders are assumed to be honest and the receivers have to receive the same messages per notion definition.

\paragraph{Allowed number of challenge rows}
The \Trilemma\footnote{Even though the \Trilemma~\cite{das2018anonymity}'s privacy notion formally allows just one communication to differ, its attack is, in combination with the assumption that every user sends exactly one message, not possible with only one differing communication.} and \Ando\ need only two differing communications (the ones whose senders are switched) in the compared scenarios. Thus, the notions of the \Trilemma\ and \Ando\ are also in this regard weaker than the one of the \Gelernter\ where multiple (precisely $\mu_{max}$) differing communications are needed for the attack.

Further, the protocol model of the \Trilemma\ allows only one user to send a real message per round and this permutation over the users is assumed to be chosen randomly. This fits our understanding of batches: The order, in which the chosen communications are input to the protocol model, is random.


\begin{table*}[h]
\caption{Comparison of notations, N.A.: not applicable (concept does not exist/does not apply), -: no defined symbol}
\label{table: notations all }
	\center
	\resizebox{0.76\textwidth}{!}{%
\begin{tabular}{lllll}
	\hline
	Our & \Trilemma & \Gelernter & \Hevia & \Ando \\
	\hline
	$\mathcal{U}_H$ & N.A. & $H$ & N.A. & -\\
	$h$ & - & $h$ & N.A. & - \\
	$c_a$ & N.A. & N.A. & N.A. & $\kappa N$\\
	$c_p$ & $c$  & N.A. & N.A & N.A. \\
	$l_{max}$ & $l+1$ & N.A & N.A. & N.A.\\
	$l_{exp}$ & N.A. & N.A. & N.A. & N.A.\\
	$r$ & N.A. & $R$ & N.A. & N.A\\
	$\sigma_0$, $\sigma_1$ & N.A. &	$\sigma_0$, $\sigma_1$ & $\bar{M}^{(0)}$, $\bar{M}^{(1)}$& $\sigma^0$, $\sigma^1$\\
	$\mathcal{U}$, $u_i$ & $\mathcal{S}$, $u_i$ & $[n], p_i$ or $S_i$ & $P_i$ & -\\
	$\delta$ & $\delta$  & $Adv^{Comp-\mathbb{N}}_{\pi,n,A,Cap}(k)$ & $Adv^{\mathbb{N}-annon}_{\pi,A}(k)$ & $\mathsf{Adv}^{\Pi, A}(\sum, \mathsf{kickoff}, \mathsf{freeze})$\\
	$\lambda$ & $\eta$ ($\delta\leq neg(\eta)$) & $k$ ($Adv$ $\leq negl(k)$) & $k$ & $\lambda$\\
	$p$ & $p=p'+\beta$&  $\approx \frac{L^\pi_i(\sigma, R)}{R}$ & $\sum_{j\in [n]}|m_{i,j}|$ & N.A.\\
	$\beta$ & $\beta$ & $\frac{Com^\pi_{\sigma,R}-Out^\pi_{\sigma, R}}{N*R}$ & $l_i= \mu_\mathbb{N} - \sum_{j\in [n]}|m_{i,j}|$ & N.A.\\
	& & &\hspace{0.35cm}$\approx \mathtt{ovh}(T) / n = \mu_\mathbb{N}$ &\\ 
	$b$ & $b$ & $b$ & $b$ & $b$\\
	$n$ & $N$ & $n$ & $n$ & $N$\\
	$Out(r)$& $\approx p' \cdot r  \cdot N $   & $Out^\pi_{\sigma, R}$ &  - & - \\
	\hline
\end{tabular}}
\end{table*}

\begin{table}[tbh]
	\center
		\caption{Overview used parameters}
	\label{table: our notations}
	\resizebox{0.46\textwidth}{!}{
		\begin{tabular}{c p{6.5cm}}
			\hline
			\textbf{Parameter} & \textbf{Meaning} \\
			\hline
			$n$ & number of nodes / participants \\
			$\mathcal{U}_H$ & set of honest senders \\
			$h$ &  number of honest senders, $|\mathcal{U}_H|$\\
			$c_a$  &  number of actively compromised nodes\\
			$c_p$ & number of passively compromised (intermediate) nodes\\
			$\mathcal{U}$ & set of senders\\
			$u_i$ & $i^{th}$ sender \\
			$m$ & message \\
			$\lambda$ & security parameter\\
			$\delta$ & adversary advantage \\
			$l_{max}$ & latency, maximal delay of a message, the maximal number of rounds between the sending of a message and its reception \\
			$l_{exp}$ & expected delay of a message, average number of rounds between the sending of a message and its reception \\ 
			 $r$ & number of rounds a certain metric or analysis refers to \\
			$\beta$ & bandwidth-overhead, \\
			& probability of one node to send a dummy message in a given round\\
			$p'$ & probability of one node to send a real message in a given round\\
			$p$ & $p=p'+\beta$, probability to send any type of message\\
			$Out(r)$ & delivered messages until round $r$\\
			\hline 
	\end{tabular}}

\end{table}

\section{Tightening the claims}\label{sec:worse}
First, we make the effects of assumptions explicit by incorporating them into the analyzed dimensions. Thereby, we do not technically change any result, but allow to understand the real strength of the results better. Secondly, by in depth analysis of the proofs, we found that the proofs work for even weaker assumptions than those that had been made, and in one case, we improved the calculations for the overhead. 

\subsection{Adversary Models}\label{sec:worseAdv}

All papers assume global eavesdropping capabilities. However, as the actual attacks consider only one or two victim senders, we can reduce the global adversary to be local. She controls the links\footnote{This can be achieved trivially by their ISP, and probably easily by an attacking insider, who controls the nodes that are connected by the adjacent links.} of the victim(s).

\subsubsection*{\Hevia, \Gelernter\ and \Trilemma} For these attacks the adversary only has to be able to notice when or how often the victim sends. In the integrated system model she has thus to be able to distinguish sending events from forwarding events. As a technicality the adversary in the proofs can decide that the victims do not receive any message. As thus all inbound packets must be followed by forwarding events, the adversary learns the number of real sending events by subtracting the outgoing packets from the inbound. 

\subsubsection*{\Ando} For the bound the only active adversarial capacity needed  is dropping, although their attacker model states multiple active capabilities (delay, create, modify and drop messages).


\subsection{\Ando\ -- Privacy Notion}
The \Ando\ defines its own privacy goal without relation to other work. In its anonymity definition the game adversary is\footnote{except for the behavior of corrupted users that  does not hinder our comparison as we discuss separately in Appendix~\ref{app:additionalRestrictions}} not restricted in how she chooses the scenarios. Therefore, the described notion matches communication unobservability $C\overline{O}$, the strongest notion in the hierarchy. 

Additionally to the anonymity definition the goal is however restricted by the "simple I/O setting", i.e. each participant sends and receives exactly one message. This restriction is equivalent to fixing the number of sending and receiving events to 1, which is the exact definition of $M\bar{O}[M\bar{L}]_1$ (see Appendix \ref{app:additional}), an already weaker impartial notion of the hierarchy. 

The attack used in the proof breaks for an even weaker notion. As it ignores message contents, we can use the same message in all communications. Further, we can define the second scenario equal to the first, with only the one sender $u_0$ that sends to the observed receiver  switched with the alternative sender $u_1$ that sends to another receiver. Thereby, only the linking between those senders and receivers differs and the definition of $(SR)\bar{L}_1$ is met.

\begin{figure}[h!]
	\center
	\includegraphics{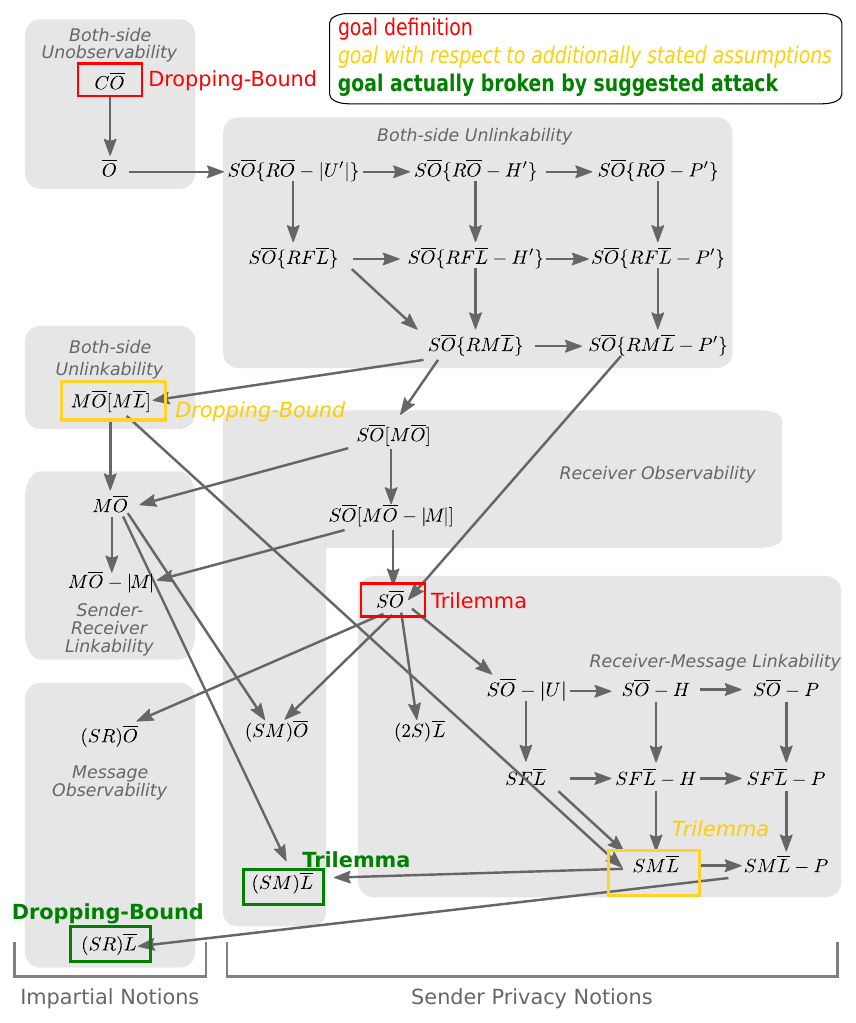}
	\caption{The mapping of the anonymity notions to the hierarchy of privacy notions from \cite{kuhn2018privacy}: The notions used for the bound differ from the anonymity definition given due to additional assumptions. Further, the notions needed in the proofs are even weaker than the notions that follow from the additional assumptions. For a simplified summarizing presentation, we neglect the additionally restriction ($X_1$) that the \Trilemma\ and the \Ando\ introduce for the notions $MO[M\overline{L}], SM\overline{L}, (SM)\overline{L}$ and $(SR)\overline{L}$.}
	\label{fig:hierachy}
\end{figure}

Interestingly, this is one of the weakest notions in the hierarchy. The bound is thus much stronger than the anonymity definition suggested (the strongest notion in the hierarchy), as their calculated cost is not only necessary to achieve a very strong privacy definition, but also if only the linking between sender and receiver is aimed to be protected (see Figure~\ref{fig:hierachy}).

\subsection{\Trilemma} \label{sec:worseTrilemma}
We discuss the used privacy notion and tighten the bound in terms of needed overhead.

\subsubsection*{Privacy Notion}

The \Trilemma\ uses sender anonymity from AnoA, which maps directly to $S\overline{O}$. This means that the two scenarios can arbitrarily differ in the senders, but in nothing else. 

For the synchronous user distribution it is however additionally assumed that everyone can only send exactly one real message. Similarly to the \Ando\ before, this means that the frequency with which a sender sends needs not to be hidden, as it is identical in both scenarios. This is equal to the definition of $SM\overline{L}_1$ (see Appendix \ref{app:additional}).

The analyzed notion in the proof, as opposed to the claimed goal or the goal that follows from the stated assumptions, of the synchronized model changes however only the sender of the challenge message (with some other sender\footnote{This is not made explicit, but has to be done to respect the assumption of every sender sending exactly one message.}). This matches the definition of $(SM)\bar{L}_1$  as  only the connection between two senders and messages is changed and every sender sends one message.


In the unsynchronized model, it is assumed that each user wants to send messages. This time the number of messages is not fixed by the notion, but neither can it be chosen by the adversary. Every time the coin flip decides that the user has to send a message, she is assumed to have one ready to send. As again the only difference allowed is the change of senders for the challenge message, this translates to $(SM)\bar{L}$. 

Similarly to the \Ando\ we see that even though a pretty strong notion was stated in the beginning, the suggested attack breaks one of the weakest notions defined; the notion that only protects the linking between sender and message, but keeps anything else identical  (see Figure~\ref{fig:hierachy}).

\subsubsection*{Bound}
The idea of the proof in the unsynchronized case is simple: An adversary knows that the sender of the critical message has sent in the $l_{max}-1$ rounds before she received this message. Thus, if one of the two victim senders did not sent in these rounds, we know the other must have been the sender, as the only uncertainty the adversary has left is which of those two candidate users was the sender. Therefore, the adversary wins, i.e. learns the sender, if the alternative sender did not send.

The authors perform intricate calculations, introducing random variables, the Chernoff bound and Markov's inequality, to prove their bound:
{\small
\begin{equation*}
\delta \geq 1-\left[\frac{1}{2} + min\left(\frac{1}{2},1-(1-p)^{l_{max}-1}\right)\right]
\end{equation*}
}
Which is equivalent to
{\small
\[\delta \geq \frac{1}{2} - min\left(\frac{1}{2},1-(1-p)^{l_{max}-1}\right)\]
}
and can be even easier understood as:
{\small
\[\delta \geq \max \left(0,(1-p)^{l_{max}-1}- \frac{1}{2}\right)\]
}

However, considering that we only need to bound the probability that the other user does not send, we claim that an easier and more accurate bound is:
\[\delta \geq (1-p)^{l_{max}-1}\]

We know that with probability $1-p$ the alternative user does not send in one round. As the sending in the rounds are independent (as stated in \cite{das2018anonymity}), $(1-p)^{l_{max}-1}$ is the probability that the user does not send in any of the rounds. 

With this difference, we adapt the argumentation of \cite{das2018anonymity} (which we explain intuitively together with the other bounds in Section~\ref{sec:mainBounds}) for the extended case with compromised protocol parties as well and result in (cf.  Appendix~\ref{app:TrilemmaImpovementComplex}): 
\begin{equation*}
\scriptsize
\delta \geq \begin{cases}
1-\left[1-\binom{c}{l_{max}-1}/\binom{K}{l_{max}-1}\right]&\left[1-(1-p)^{l_{max}-1}\right] \\&  \text{if } c_p\geq l_{max}-1\\
\left(1-\left[1-1/\binom{K}{c_p}\right] \left[1-(1-p)^{c_p} \right] \right)& \left( (1-p)^{l_{max}-1-c_p} \right) \\& \text{if } c_p<l_{max}-1\\
\end{cases}
\end{equation*}

\section{Delayed Comparisons}
\subsection{\Hevia\ vs. \Gelernter} \label{app:HeviaGelernter}
The \Hevia \cite{hevia2008indistinguishability}: Hevia and Miccianchio investigated performance optimality of ACN protocol transformations against a global passive adversary. A transformation is a technique that can be added to a protocol achieving a weak privacy goal to create a stronger protocol. 
They prove a transformation optimal; thus stating a performance bound. 

\subsubsection{Privacy Goal}
They use the same definition of protected communication properties and as stated above the advantage definition is equivalent.

\subsubsection{Bound}

The \Hevia proves optimality of a protocol transformation: The overhead $\mathtt{ovh}$ of each such protocol transformation $\tau$ has to ensure that each of the $n$ possible senders is sending the maximum number of messages $\mu_{max}$. This leads to $n \cdot \mu_{max}$ send events:
\begin{equation*}
\mathtt{ovh}(\tau)\geq  n \cdot  \mu_{max}
\end{equation*}

As $\mu_{max}$ messages are delivered $Out(r) = \mu_{max}$. The total number of messages sent are $Com(r)= ovh(\tau)$. Since there are no corrupted users $h=n$. From this we conclude equality to the \Gelernter : 
\[ \mathtt{ovh(\tau)}\geq n \cdot  \mu_{max} \iff Com(r) \geq n \cdot Out(r) = h \cdot Out(r)  \]\\

\subsection{Trilemma's Compromising adversary} \label{App:TrilemmaCompromising}
Extending the adversary to compromise up to $c_p\leq n-2$ intermediate nodes facilitates the attack of tracing messages along their anonymization paths, if all nodes on these paths are under adversarial control.
This increases the advantage of the adversary, and the \Trilemma\  is interested in this additional probability for an attack to succeed.
We use $K$ to denote the number of protocol parties throughout this section and discuss the \emph{synchronized user setting} in the following. 

Recall that the bound for synchronized users without corruption is:
\[\delta \geq 1-min\left(1,\frac{(l_{max}-1)(1+\beta n)}{n-1}\right)\]

According to \cite{das2018anonymity} we
define the last part (the probability that a certain user has sent a message in the $l_{\max}-1$ rounds) to be \[f_{\beta}(l_{\max} -1):= min\left(1,\frac{(l_{max}-1)(1+\beta n)}{n-1}\right)\text{.}\]

For corrupted intermediaries, \cite{das2018anonymity} distinguishes two cases.
The adversary either has a chance to compromise all relays on the anonymization path of the challenge message as she has corrupted enough relays, or not.
The authors simplify the first case and bound the probability that the challenge or alternative messages can be traced with the probability that all relays on the anonymization path are compromised:
$\binom{c_p}{l_{max}-1}/\binom{K}{l_{max}-1}$. 
The adversary can only lose if  some relay  on the path is honest $\left(1-\frac{\binom{c_p}{l_{max}-1}}{ \binom{K}{l_{max}-1}}\right)$ and an alternative message is sent ($f_{\beta}(l_{\max} -1)$). She thus loses with a probability of at most $\left(1-\frac{\binom{c_p}{l_{max}-1}}{ \binom{K}{l_{max}-1}}\right) f_{\beta}(l_{\max} -1)$. As she wins  in the complement to this event, her advantage in this case is at least:
\[ 1-\left(1-\frac{\binom{c_p}{l_{max}-1}}{ \binom{K}{l_{max}-1}}\right) f_{\beta}(l_{\max} -1).   \]

In the second case, not all intermediate nodes can be corrupted.
Note that for the adversary to win it suffices to track all alternative messages until the challenge message is received (as she can exclude them).
 The adversary hence loses if an alternative message is sent and an honest relay is on the path that this message shares with the challenge message. There is an honest relay on this path if the message traversed more relays ($>\!c_p$) than the adversary can compromise ($f_{\beta}(l_{\max} -1 - c_p)$). However, there might also be an honest relay on this path if the path is shorter (consisting of $\leq\! c_p$ relays). This event is at most as likely as having an honest relay in exactly $c_p$ relays: $1-1/\binom{K}{c_p}$. As a shorter path occurs with probability $f_{\beta}(c_p)$, the adversary loses at most with the probability $f_{\beta}(l_{\max} -1 - c_p) + f_{\beta}(c_p)(1-1/\binom{K}{c_p})$. The adversary wins in the complementary event, so her advantage is at least
\[1-\left[1-1/\binom{K}{c_p}\right]f_{\beta}(c_p) - f_{\beta}(l_{\max} -1 - c_p)\text{.}\]

The two considerations result in the final bound: 
\begin{equation*}
\tiny
\delta \geq \begin{cases}
 1-\left[1-\binom{c_p}{l_{max}-1}/\binom{K}{l_{max}-1}\right]f_{\beta}(l_{\max} -1)&  c_p\geq l_{max}-1\\
  1-\left[1-1/\binom{K}{c_p}\right]f_{\beta}(c_p) - f_{\beta}(l_{\max} -1 - c_p)&  c_p< l_{max}-1
  \end{cases}
\end{equation*}

For the \emph{unsynchronized setting} the same ideas are applied on the basis of the non-compromising bound for the unsynchronized setting (cf. Appendix~\ref{app:TrilemmaImpovementComplex}).

As for the non-compromising case, the above bounds induce an \emph{area of impossibility for the compromising adversary}. 
If an adversary passively compromises  $c_p<l_{max}-1$ protocol parties, then the area of impossibility is  \[2(l_{max}-1-c_p)\beta \leq 1-\frac{1}{poly(\lambda)}\text{.}\]
If the number of compromised nodes is ${c_p\geq l_{max}-1}$, then anonymi\-ty cannot be reached for  \[2(l_{max}-1)\beta \leq 1-\frac{1}{poly(\lambda)} \text{ and } l_{max}\in O(1)\text{.}\]

\section{Proofs}
\subsection{Improving the \Trilemma} \label{app:TrilemmaImpovementComplex}
\paragraph*{Case 1: $c_p\geq l_{max}-1$}
This means all intermediate nodes chosen in the $l_{\max}-1$ rounds could be corrupted.
As for the synchronous behavior, the attackers definitively wins if all intermediate nodes are corrupted ($\binom{c}{l_{max}-1}/\binom{K}{l_{max}-1}$). He also wins if the alternative user does not sent ($(1-p)^{l_{max}-1}$). So, her advantage can be bound by the complementary event to not all intermediate nodes being corrupted ($1-\binom{c}{l_{max}-1}/\binom{K}{l_{max}-1}$) and the probability that the other user sends ($1-(1-p)^{l_{max}-1}$):
{\small
\[ 1-\left[1-\binom{c}{l_{max}-1}/\binom{K}{l_{max}-1}\right]\left[1-(1-p)^{l_{max}-1}\right] \]
}
\paragraph*{Case 2: $c_p<l_{max}-1$}
This means not all intermediate nodes are corrupted.
As for the synchronous behavior, the attacker wins except if an alternative and the challenge message share long path (so long that an honest node has to be on it) ($ 1- (1-p)^{l_{max}-1-c_p} $) or there are \emph{only} alternative messages that share short paths (and none that shares a long path)\footnote{Note that this is a tighter estimation as the one of synchronized user setting, where the probability of  a short shared path($f_{\beta}(c_p)$) is used (and the existence of further alternative messages is neglected).} ($ (1-p)^{l_{max}-1-c_p}(1-(1-p)^{c_p})$) but an honest node is on it($\leq 1-1/\binom{K}{c_p}$):
{\tiny
\begin{align*}
\delta &\geq 1-\left( 1- (1-p)^{l_{max}-1-c_p} \right)\\& \qquad  - (1-p)^{l_{max}-1-c_p}\left[1-(1-p)^{c_p}\right]\left[1-1/\binom{K}{c_p}\right] \\
&=(1-p)^{l_{max}-1-c_p}  - (1-p)^{l_{max}-1-c_p}\left[1-(1-p)^{c_p}\right]\left[1-1/\binom{K}{c_p}\right]\\
&=\left((1-p)^{l_{max}-1-c_p}\right)  \left( 1- \left[1-(1-p)^{c_p}\right] \left[1-1/\binom{K}{c_p}\right] \right)\\
\end{align*}
}%

\subsection{Impossibility areas}\label{appendix: gelernter and trilemma: equations}
%

\paragraph*{Relations between variables}
The number of send messages $Com(r)$ are the dummy messages ($\beta$ messages per user and round ) and real messages that are delivered ($Out(r)$).
\[Com(r)  = \beta n r + Out(r)\]

\paragraph*{Transformation}
Using our discovered relation between the variables and the Trilemma's assumption that $n\approx poly(\lambda)$ we can transform the \Gelernter:
\begin{align*}
Com(r) &\geq Out(r)\cdot n \\
\beta n r + Out(r) &\geq Out(r) \cdot poly(\lambda)\nonumber\\
\beta &\geq \frac{Out(r) \cdot (poly(\lambda)-1)}{poly(\lambda) \cdot r}\nonumber\\
\beta &\geq \frac{Out(r)}{r}  \cdot   \left(1-\frac{1}{poly(\lambda)}\right)
\end{align*}

and the impossibility area of the \Trilemma :
\resizebox{ \columnwidth}{!} 
{
$ 2(l_{max}-1)\beta \geq 1-\frac{1}{poly(\lambda)} \iff
\beta \geq \frac{1}{2(l_{max}-1)}\left(1-\frac{1}{poly(\lambda)}\right) $
}%

%
%

\subsection{No latency in the Trilemma} \label{app:latencyNull}
Using $l_{max}=1$ yields:

\resizebox{0.46\textwidth}{!}{%
\begin{tabular}{ll}
synchronized:& $\delta \geq 1 - min\left(1,\frac{0+\beta n 0}{n-1}\right) = 1$\\
unsynchronized (original):&$\delta \geq 1-\left[\frac{1}{2} + min\left(\frac{1}{2},1-(1-p)^{0}\right)\right]= \frac{1}{2}$\\
unsynchronized (improved):&$\delta \geq (1 - p)^0 = 1$
\end{tabular}}

 \section{Receiver Privacy Goals}\label{sec:receiver}
 Both the \Hevia\ as well as the \Trilemma\ also consider receiver privacy goals.

The analysis for the \Hevia\ is consistent to its bound for $S\bar{O}$: if any user receives less than the total number of real messages, she is excluded from the anonymity set.
Both sender and receiver bounds hence are equal, and $R\bar{O}$ can only be achieved with high bandwidth overhead; for instance, by implementing a broadcast.


The \Trilemma\ also adapts its original attack to identify receivers:
%
The adversary observes the sending of the challenge message and concludes that the message can only be received by someone who receives a message within the next $l_{max}$ rounds. 
If enough relays are corrupted, the message can be traced. 
Interestingly, the resulting bound postulates a lower cost than for the senders.
Attacking the sender, the candidate messages are only those sent within the $l_{max}$ rounds before the challenge message is received.
Attacking the receiver, the candidate set expands to those messages sent during the $l_{max}-1$ rounds before, and the $l_{max}-1$ rounds after sending the challenge message.
All could have caused a message reception during the critical period, depending on how the protocol determines the latency for each message.
We hypothesize that future work might improve this bound to match the sender case, because not even an optimal protocol can be able to ensure that all messages  \emph{always} end up being received in the critical period\footnote{As receiving all these messages in one $l_{max}$ interval implies that less receive events occurred during the $l_{max}$ interval before.}.

 \section{Note on related results}\label{sec:discussion}
%

It is interesting to note, that researchers on the physical layer defined privacy goals that are similar, and identified the same bound as the Optimality- and the Counting-Bound \cite{sun2018capacity}.
Assuming the lack of a shared secret, they additionally analyze how much shared randomness is needed between the users.


Oya et al. analyze how a given amount of dummy traffic should be spent in pool mix networks to optimally improve their privacy \cite{oya2014dummies}.
This differs in two ways from the analyses we systematized in this work:
First, their privacy measure considers a mean over all users. This is conceptually different from game-based approaches, which always consider the worst-case user or user-pair. 
Second, they do not give a bound on dummy messages required to achieve a certain privacy goal, but show how to best use a dummy traffic budget.

\end{document}